\documentclass[pdflatex,sn-mathphys-num]{sn-jnl}

\usepackage{float}
\usepackage{amsmath}
\usepackage{amssymb}
\usepackage{graphicx}
\usepackage{caption}
\usepackage{subcaption}
\usepackage{algorithm}
\usepackage{algpseudocode}
\usepackage{listings}
\usepackage{makecell}
\usepackage{url}
\usepackage{xcolor}
\usepackage{booktabs, multirow} 
\usepackage{soul,color}
\usepackage{changepage,threeparttable} 
\usepackage{svg}
\usepackage{amsfonts}%
\usepackage{amsthm}%
\usepackage{mathrsfs}%
\usepackage[title]{appendix}%
\usepackage{textcomp}%
\usepackage{manyfoot}%
\usepackage{algorithmicx}%

\soulregister\Hl{7}
\soulregister\ref7
\soulregister\cite7
\soulregister\pageref7

\theoremstyle{thmstyleone}%
\theoremstyle{thmstyletwo}%

\theoremstyle{thmstylethree}%

\begin{document}

\title[FPGA-DSP Montgomery Multiplier]{Analyzing the capabilities of HLS and RTL tools in the design of an FPGA Montgomery Multiplier}


\author[1]{\fnm{Rares} \sur{Ifrim}} \email{rares.ifrim@upb.ro} 

\author[1]{\fnm{Decebal} \sur{Popescu}} \email{decebal.popescu@upb.ro}

\affil[1]{\orgname{National University of Science and Technology POLITEHNICA Bucharest}, \orgaddress{\country{Romania}}}


\abstract{We present the analysis of various FPGA design implementations of a Montgomery Modular Multiplier, compatible with the \textit{BLS12-381} elliptic curve, using the Coarsely Integrated Operand Scanning approach of working with complete partial products on different digit sizes. The scope of the implemented designs is to achieve a high-frequency, high-throughput solution capable of computing millions of operations per second, which can provide a strong foundation for different Elliptic Curve Cryptography operations such as point addition and point multiplication. One important constraint for our designs was to only use FPGA DSP primitives for the arithmetic operations between digits employed in the CIOS algorithm as these primitives, when pipelined properly, can operate at a high frequency while also relaxing the resource consumption of FPGA LUTs and FFs. The target of the analysis is to see how different design choices and tool configurations influence the frequency, latency and resource consumption when working with the latest AMD-Xilinx tools and Alveo FPGA boards in an RTL-HLS hybrid approach. We compare three categories of designs: a Verilog naive approach where we rely on the Vivado synthesizer to automatically choose when and where to use DSPs, a Verilog optimized approach by manually instantiating the DSP primitives ourselves and a complete High-Level Synthesis approach. We also compare the FPGA implementations with an optimized software implementation of the same Montgomery multiplier written in Rust.}


\maketitle

\section{Introduction}

Optimizing the operations encountered in the Finite Fields arithmetic can bring a big improvement in the upper layers of an application stack such as an Elliptic Curve Cryptography scheme~\cite{wenger2011exploring}(for example a digital signature scheme such as the ECDSA~\cite{johnson2001elliptic}), or even in a blockchain~\cite{nakamoto2008bitcoin} or zero-knowledge proof (ZKP) system~\cite{tran2023implementation}. Research has been done in multiple directions for improving the Finite Field arithmetic operations, especially the modular multiplication (MM)~\cite{nedjah2006review}, such as optimizing the algorithms themselves when it comes to software execution on a CPU~\cite{botrel2023faster}, or offloading and accelerating them on specialized hardware structures such as a specialized CPU extension like the Intel AVX or ARM NEON~\cite{cornea2015intel,6670327} or hardware accelerators such as GPUs~\cite{emmart2016optimizing}, FPGAs~\cite{8968620} or even ASICs~\cite{10246322}.

As modular multiplication has a big impact on these applications, we selected this operation as the main subject of our analysis, where we seek to implement a fast modular multiplier unit with the scope of using it in future FPGA-based (Field Programmable Gate Array) kernels that accelerate high throughput demanding systems such as blockchains or ZKPs. The arithmetic unit is developed and evaluated on an AMD Alveo U55C platform~\cite{u55c}. The choice of using an FPGA comes from the flexibility, performance and power efficiency that the technology offers, while the Alveo family of boards from AMD is focused on highly intensive computing tasks. This particular board is based on the Virtex UltrascalePlus family architecture~\cite{ultrascaleplus} that comes with the DSP48E2 arithmetic blocks, offering support for 27x18 integer multiplications and 48x48 integer additions which we make use of in our implementations~\cite{leibson2013xilinx}.  

As the integrity, security and trust of ECC-based applications rely on the safeness of the elliptic curve used, we selected the \textit{BLS12-381} elliptic curve~\cite{cryptoeprint:2019/403}, a short Weierstrass curve, as base for our modular multiplier unit size (384 bits) and field modulus (381 bits prime) as it is a newer curve being adopted and standardized already in popular blockchain systems such as Ethereum~\cite{2020eip2537, 2019org} that came as a response after the ECC attack in CRYPTO 2016~\cite{kim2016extended}.

The scope of this analysis is the implementation and evaluation of different design choices made in the development of an FPGA-based Montgomery Modular Multiplier for the \textit{BLS12-381} curve using the latest versions of Vitis and Vivado, a popular choice when it comes to FPGA design and implementation. The implementation goal of the multiplier is to achieve a high-frequency and high-throughput unit, that is resource-aware. For this, we made our design choices with the main focus of using only DSPs (and Flip-Flop registers where needed) for the integer arithmetic operations (additions, multiplications, modulo reduction), as when pipelined properly, these FPGA primitives provide high performance, are power-efficient, and can relax the other resource usage. The algorithm selected for implementation is the Coarsely Integrated Operand Scanning (CIOS) Montgomery algorithm~\cite{502403} as it offers benefits such as interleaving of multiplication and reduction steps, replacing the division by the field modulus with right shifts and offering a more efficient chain of modular operations.

For our implementations, we choose three word sizes for the Montgomery CIOS algorithm implementation, namely 24, 32 and 64 bits as they map efficiently on DSPs primitives. We tackle the implementation by two approaches: a classic RTL (Register-Transfer Level) style of implementation based on Verilog and a more productive approach through High-Level Synthesis (HLS)~\cite{o2014xilinx}. These two approaches map further to three main categories of design implementations: (i) manually instantiating the DSP units in Verilog for better efficiency and fine-grained control over the integer operations implementations; (ii) implementing the modular multiplier units completely in HLS and leveraging the compiler's available pragmas which allow for faster time-to-market and better design exploration; and (iii) a naive Verilog implementation based on a better multiplication algorithm, but using the synthesizer's \textit{use\_dsp} flag available in Vivado to see how and if the tool can better optimize the DSP usage compared to our manual instantiations. In total we implement and evaluate 19 different designs, using the Vitis Unified Software Platform 2023.2 environment from AMD (former Xilinx).

As part of our analysis, we also explore and present an RTL-HLS hybrid approach of integrating the Montgomery Multipliers into a complete system. This is motivated by some of the success stories in the industry that showcase how such an approach can offer the best of both worlds when it comes to hardware design~\cite{hls_nvidia,hls_stm}. The hardware wrapper written in HLS uses the RTL-blackbox feature of Vitis, which allows us to easily plugin our Verilog implementations as black-boxes in a hardware kernel that can be interfaced over PCIe (Peripheral Component Interconnect Express) from a host application. Through this hardware wrapper we can easily test and benchmark all our Montgomery Multiplier designs. In summary, we analyze and compare the frequency, throughput, resource an power usage of various Montgomery Multiplier designs where we manually instantiate DSPs in Verilog, automatically infer DSP usage through synthesizer flags and use the HLS pragmas on the latest FPGAs and software tools provided by AMD-Xilinx.    

The structure of the paper is as follows. Section~\ref{sec:background} presents the Montgomery Modular Multiplication, alongside with details about the Finite Field arithmetic and the CIOS approach used for this case study. The hardware wrapper and the main details used for testing each design are depicted in Section~\ref{sec:framework}, while Section~\ref{sec:design} elaborates on the actual implementation details of each evaluated design. The results obtained are shown in Section~\ref{sec:results}. Finally, the last two sections show the current limitations of our designs, present potential enhancements and summarise our remarks and conclusions resulting from this analysis.

\section{Background}
\label{sec:background}

\subsection{Coarsely Integrated Operand Scanning Montgomery algorithm}
\label{cios_def}

Montgomery Modular Multiplication (MMM)~\cite{montgomery1985modular} speeds up the modular multiplication operation $t = a\times b\mod p$ by converting the two operands into the so-called Montgomery domain, with the advantage of avoiding the division by the modulus $p$. The conversion requires a large integer $R$ such that $R >> p$ and $gcd(R,p) = 1$. For $p$ odd and size of $p$ being $n$ bits, we can choose $R = 2^n$ so that division can be easily replaced with cheap left shift operations. In the Montgomery domain, the modular reduction and division operations are performed on $R$ instead of $p$, and $p$ is involved only in integer multiplications (see Algorithm~\ref{mmm_alg}).

For converting an integer $a$ to the Montgomery domain counter-part $\bar{a}$, we define it's $p$-residue with respect to $R$ and perform a modular multiplication as:
\begin{equation}
\bar{a} = a \times R \mod p
\end{equation}

Performing addition and subtraction is straight-forward as adding/subtracting the two numbers directly in the Montgomery domain:

\begin{equation}
\begin{gathered}
\overline{sum} = \bar{a} + \bar{b} \mod p = (a + b) \times R \mod p \\
\overline{sub} = \bar{a} - \bar{b} \mod p = (a - b) \times R \mod p 
\end{gathered}
\end{equation}

Converting the results back from the Montgomery domain implies performing another modular multiplication with the inverse of $R$ defined by $R^{-1}\times R = 1 \mod p$:

\begin{equation}
a = \bar{a} \times R^{-1} \mod p
\end{equation}

Multiplication in the Montgomery domain requires an extra multiplication by the same inverse $R^{-1}$ motivated by Equation~\ref{monprod} where we consider that $prod = a \times b$ in our original field domain:

\begin{equation} \label{monprod}
\begin{gathered}
\overline{prod} = \bar{a} \times \bar{b} \times R^{-1} \mod p \\  
\overline{prod} = a \times R \times b \times R \times R^{-1} \mod p \\
\overline{prod} = a \times b \times R \mod p \\
\overline{prod} = prod \times R \mod p \\
\end{gathered} 
\end{equation}

Without taking into consideration the conversion steps to and from the Montgomery domain, going by the naive approach of Equation~\ref{monprod} is still slower than multiplication in the standard field because of the multiplication by $R^{-1}$ (which is not a power of 2 anymore). On top of this, we still need to perform the complex $mod\;p$ operation as in the first case. A better approach is using the Montgomery reduction performed between the second and last line of Algorithm~\ref{mmm_alg} which multiplies the product of $a$ and $b$ by $p'$ where $p \times p' + R\times R^{-1} = 1$ in order to avoid division. In other words, we can say that there is a $p'$ which satisfies the equation $R \times R^{-1} = 1 \mod p \Longleftrightarrow R \times R^{-1} = p \times p' + 1$. This allows us to use $R$ in the modular and division steps instead of $p$ which is much faster as it is a power of two. Checking that the final result is within the boundaries of $[0,p-1]$ means only performing a simple subtraction of $p$ in the case of an overflow. In a real-world scenario, $R$, $R^{-1}$ and $p'$ can be pre-computed for a known $p$ and size of $p$ in order to avoid extra computations for each pair of given inputs.

\begin{algorithm}[!htp]
\caption{Montgomery Modular Multiplication}\label{mmm_alg}
\begin{algorithmic}
\renewcommand{\algorithmicrequire}{\textbf{Input:}}
\Require $a, b$ in Montgomery domain,\par
$p$ odd field modulus,\par
$R$ larger than $p$ and co-prime with it
\State $t = a \times b \mod R$
\State $m = t \times p' \mod R$
\State $u = t + (m \times p) / R$
\If{$u \geq p$}
    \State Return $u - p$
\EndIf
\State Return $u$
\end{algorithmic}
\end{algorithm}

Taking into account the need to convert the operands to and back from the Montgomery domain, multiplying them, and the intermediate multiplications with $p$ and $p'$ the Montgomery Multiplication algorithm is not efficient when performing a single operation, but becomes so when there is a chain of modular operations applied to the same operands as we see in the case of a point addition algorithm~\cite{meloni2007new}. Thus, in a use-case like this, it makes sense to convert the coordinates to the Montgomery domain, apply the chain operations of modular multiplications and modular additions/subtractions, and then convert them back, the overhead of this being just two extra modular multiplications for the conversion steps.

Public-key applications like ECC and RSA use large integer fields (over 100 bits) to harden the discrete-logarithm problem~\cite{sp2023recommendations}. Computing these finite field operations on classic processors or reconfigurable devices like FPGAs poses problems as CPUs use smaller word sizes (32 or 64 bits), and in the case of FPGAs, even though it provides flexibility in developing the desired datapath size that can handle these kind of integers, implementing units and buses that operate directly on these sizes increases the fan-out, creates congestion in the routing and ultimately affects the performance and area of the final design. 

As cryptographic applications like RSA and ECC gained more and more popularity, this motivated the research and development of more high-speed and space-efficient algorithms~\cite{5470228}. The method used by these algorithms is to break these large integers into smaller $w$-bit words which in turn can be better handled by the architecture of a $w$-bit processor. As it turns out, this approach is also suitable for FPGAs as it allows the synthesis and implementation tools to better place and route the needed processing elements. Some approaches further involve processing of these $w$-bit words in either serial or parallel manner or even both. The work of \cite{502403} provides us a better view of such algorithms like the Coarsely Integrated Operand Scanning (CIOS) approach for the Montgomery Multiplication and others similar to enumerate a few: FIOS - Finely Integrated Operand Scanning, SOS - Separated Operand Scanning, etc. 

There are two characteristics that describe the algorithms from \cite{502403}. The separation or integration of the reduction and multiplication steps is the first one. In the divided method, the algorithm multiplies the operands $a$ and $b$ before carrying out a Montgomery reduction. The integrated approach, on the other hand, alternates between multiplication and reduction. Depending on how frequently multiplication and reduction are switched between, the integration can be either coarse or fine-grained. This transition can happen after processing a single word or a series of words. 

The general form of the steps in multiplication and reduction is the second characteristic. One method is operand scanning, in which each word of one of the operands is traversed by an outer loop while an inner loop iterates over the words of the second operand. Product scanning is the other available method in which the loop navigates through the product's words instead. It makes no difference to these scanning techniques whether the steps for multiplication and reduction are integrated or separated. Even with the integrated approach, multiplication can also take one form while reduction can take another.

For the case of the \textit{BLS12-381} curve, as the prime field size is 381-bits, we define the big integers over 384-bits as it offers better mapping on common processor word sizes such as 32 and 64 bits. For our study case we analyze these two sizes, plus the 24bit word size which also fits perfectly into 384-bit numbers and also provides a good mapping into DSPs. While 16bit words also offer an even greater fit for the use of DSPs, this would come with greater latency as we would need to process 24 words for each operand (through the outer and inner loops) as shown in Algorithm~\ref{cios_mmm}. Because of this, we decided to not include the 16-bit word size in our analysis, and also because of inefficient mapping into DSPs, we also didn't go further up from 64-bits.

Algorithm~\ref{cios_mmm} shows a simple, non-optimized implementation of the CIOS approach for the Montgomery multiplication for the case of splitting the 384-bit inputs into  $s$ $w$-bit words. The outer loop scans over operand $b$, while the two inner loops that perform the interleaved multiplication and reduction steps, scan over operand $a$. Between each multiplication and reduction, we compute the new $m$ based on the pre-computed $p'$ and the current least significant word of the product performed in the first inner loop. All the partial products are saved into a 2$s$ word array $t$ and during each outer loop iteration, the products are shifted one by one into the upper positions of the array. At the end of the computation, the final result is to be found in the upper $s$ words of our $t$ array (during each reduction step $i$ each lower position $t[i]$ becomes $0$).

\begin{algorithm}[!htp]
\caption{Coarsely Integrated Operand Scanning Montgomery Multiplication}\label{cios_mmm}
\begin{algorithmic}
\renewcommand{\algorithmicrequire}{\textbf{Input:}}
\Require $a, b$ as arrays of $s$ $w$-bit words in Montgomery domain, \par
$p$ field modulus as an array of $s$ $w$-bit words,
\Ensure $p'$ pre-calculated as a single $w$-bit word
\State Let $t = [0, 0, \cdots ,0]$ be an array of size 2$s$ $w$-bit words
 \For{\texttt{$i \gets 0$ to $s$}} \Comment{Perform operand scanning}
    \State Let $carry = 0$ be a $w$-bit word  
    \For{\texttt{$j \gets 0$ to $s$}} \Comment{Execute multiplication step}
        \State $(carry, t[i+j]) = a[j] \times b[i] + t[i+j] + carry$ \Comment{Result is a 2$w$-bit word}
    \EndFor
    \State $t[i+s] = t[i+s] + carry$ \Comment{Propagate extra $carry$ to next word}
    \State Let $m = t[i] \times p'$ be a $w$-bit word \Comment{Prepare $m$ for reduction step}
    \State $carry = 0$ \Comment{Reset $carry$ for reduction step}
    \For{\texttt{$j \gets 0$ to $s$}} \Comment{Execute reduction step}
        \State $(carry, t[i+j]) = p[j] \times m + t[i+j] + carry$ \Comment{Result is a 2$w$-bit word}
    \EndFor
    \State $t[i+s] = t[i+s] + carry$ \Comment{Propagate extra $carry$ to next word}
\EndFor
\State Return $\{t[2s-1], t[2s-2], \cdots , t[s]\}$ \Comment{Final result is stored in last $s$ words of $t$}
\end{algorithmic}
\end{algorithm}

\subsection{Current approaches in FPGA hardware design}

While HDLs~(Hardware Description Languages) like VHDL and Verilog still dominate the market in terms of preferred languages for digital design, a report from 2022~\cite{fpga_trends} shows that in the last years, C/C++ saw an ascending adaption across multiple market segments that involve digital design. As of 2022, the adoption percentage of C/C++ was the same as the one of SystemVerilog for FPGA-based projects and was similar to VHDL for ASIC-based projects~\cite{asic_trends}.

High-level synthesis (HLS) is a groundbreaking approach to custom hardware design that serves as a connection between the familiar world of software and the intricate realm of hardware. Unlike traditional RTL coding, HLS enables engineers to express their designs using high-level languages such as C/C++ or SystemC, significantly simplifying the development process. This abstraction not only reduces design complexity but also opens up hardware design to a wider range of engineers with primarily software expertise, democratizing the field.~\cite{nane2015survey}.

While high-level programming tools have improved accessibility, there is still a learning curve for many software developers. The field lacks standardized interfaces and libraries, hindering the portability and reusability of FPGA accelerators. Efforts to seamlessly integrate FPGAs with conventional computing systems are ongoing, intending to make FPGA acceleration more accessible to a wider range of applications~\cite{5209959}.

Motivated by these findings we want to see the Quality of Results (QoR) that a tool like Vitis HLS can bring when it comes to designing a small but critical component such as a finite field multiplier with the target of having a good throughput while being area-aware. As the CIOS algorithm is suited for languages like C++, we care to see if the Vitis HLS compiler can translate this algorithm into an efficient design with minimal intervention from our side, apart from using specific compiler pragmas~\cite{vitis_hls} to guide the compiler into using DSP units, pipelining the design or partitioning a memory array.

At the same time, Verilog is still a key player in digital design. We want to see if the latest Vivado synthesizer can efficiently use the available resources of the FPGA. This can be achieved through its automatic behavior when specifying a performance-oriented synthesis strategy or by providing minimal guidance, such as using synthesizer flags like \textit{use\_dsp}. For this use case, we implement a more efficient multiplication algorithm (the Karatsuba algorithm~\cite{eyupoglu2015performance}) while using a more naive approach to describing the hardware in Verilog and relying on the synthesizer to do the heavy lifting

\subsection{Ultrascale Architecture DSP Slice}
\label{dsp_sec}

The DSP slices available in the Virtex UltrascalePlus architecture we use for evaluating our designs offer over 40 dynamically controlled operations such as pre-adder squaring, wide XOR, 27x18 multiplier, multiplier-accumulator (MACC), etc. These logic elements, called DSP48E2 (updated version from the previous DSP48E1), are capable of operating at frequencies of up to 891MHz, and some of the largest Virtex UltrascalePlus FPGA platforms contain nearly 12,000 DSP slices organized and cascaded in columns for fast chain operations (column configurations allow cascading up to 120 consecutive DSPs through the cascade-out and cascade-in ports)~\cite{ultrascaleplus}.

\begin{figure}[!htp]
	\centering
    \includegraphics[width=1\textwidth]{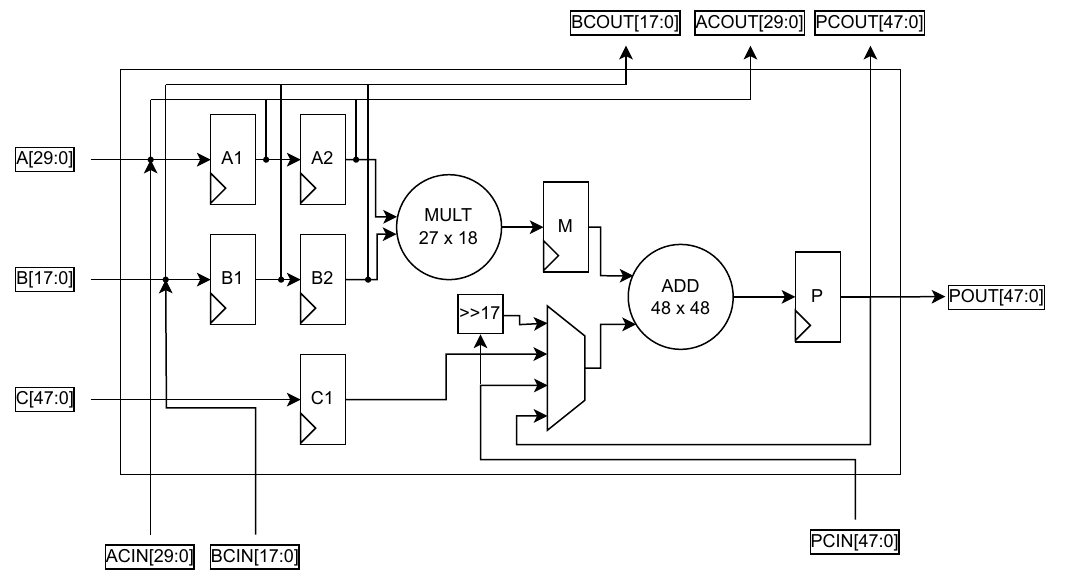}
    \caption{DSP48E2 slice high-level view of internal registers, units and ports}
    \label{fig:dsp48e2}
\end{figure}

Figure~\ref{fig:dsp48e2} depicts the high-level architecture of the DSP48E2 present in the Ultrascale FPGA family. The figure is simplified to show the main elements of the DSP logic element which we use in the implementation of our designs. It has an asymmetrically signed multiplier, capable of performing multiplications up to 27x18, and also has a 48-bit adder which can be configured to also perform two parallel 24-bit additions or four 12-bit additions for SIMD (Single Instruction Multiple Data) applications. The DSP unit also contains a 27-bit pre-adder, before the multiplier, which is omitted in this figure.

As the FPGA architecture is a column-based array, the DSP unit comes with cascade-in (CIN) and cascade-out (COUT) ports for the A, B and POUT which are connected to the preceding and succeeding DSP units inside the same column. The A, B, C, D and POUT ports are routed to left and right neighbor columns that contain fabric logic (such as LUTs, FFs) inside CLBs (Custom Logic Blocks).

Based on the needed latency and frequency, the DSP unit can be configured to work with different pipeline stages, or no pipeline at all, through the internal registers A1, A2, B1, B2, C1, M, and P. These registers can be bypassed in all combinations possible through the use of internal multiplexers and register-enable signals, and can help create appropriate delays for implementing larger adders and multipliers. The ACOUT cascade-out port can be configured to cascade either A or ACIN directly, or either of the pipeline registers A1 or A2 (similar behavior is available for BCOUT). By cascading A and B through the DSP slice, one avoids having to route the same A or B signal through fabric logic and can remain inside the DSP column where maximum performance is ensured.

In order to achieve maximum frequency, proper pipelining should be employed, meaning that all inputs, outputs and intermediate results (such as the multiplier) should be registered. Depending on the operation used, different number of pipeline stages can be enabled, the maximum depth of a fully pipelined DSP unit being four. In the case where one-step operations are executed (for example just using the multiplier, or just using the 48bit adder), it is enough to use two pipeline levels, one stage for registering the inputs and one stage for registering the outputs, which allows one to achieve the maximum possible frequency of the logical element. The internal units of the DSP, the adder, multiplier, pre-adder, etc and the corresponding registers can be disabled dynamically when not used in order to provide a low-power consumption.  

These units can be manually instantiated or inferred through behavioral description in Verilog, but can also be automatically placed by the synthesizer or HLS compiler through the use of synthesizer flags or HLS pragmas. For our analysis, we seek to find out if today's synthesis and HLS compilation tools like Vivado and Vitis can obtain similar or better results compared to the former case of instantiation or inference.

\section{HLS hardware wrapper and RTL blackboxes}
\label{sec:framework}

In order to use the same benchmark host application across all our designs, we implemented a fast HLS wrapper which allows us to easily interface with the host server through PCIe and be able to send and read our test input vectors and output results at maximum bandwidth. For this, we used the out-of-the-box XRT (Xilinx Runtime Library)~\cite{xrt}, an open-source software interface that offers an abstraction for communication between the application code and the accelerated kernels deployed on the reconfigurable portion of the PCIe-based Alveo FPGA. The Alveo U55C FPGA used for testing has installed a corresponding static shell, that comes with a direct memory access (DMA) subsystem (called XDMA~\cite{xdma}) that facilitates the transfer of data from PCIe to the available global memory of the device.

\begin{figure}[!htp]
\centering
    \includegraphics[width=1\textwidth]{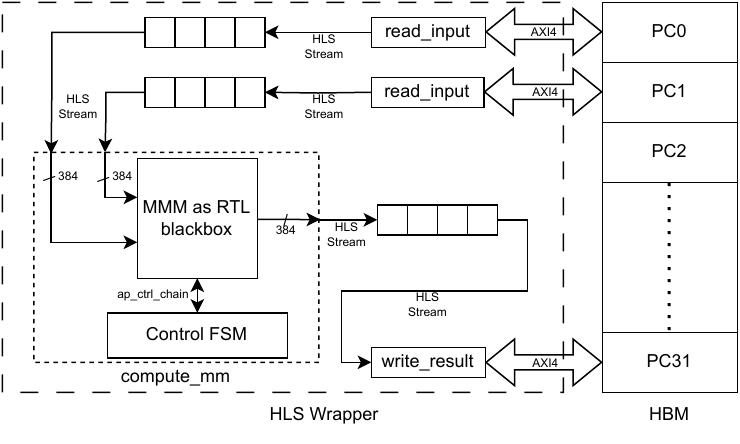}
    \caption{First variant of the HLS wrapper, the unified design containing all modules for the reading, computing and writing tasks}
    \label{fig:hls_wrapper1}
\end{figure}

The HLS wrapper comes written in two variants. The first one, depicted in Figure \ref{fig:hls_wrapper1} is implemented as a single kernel object, in a dataflow manner, where data flows from a producer-type function or module to a consumer-type one, through HLS streaming channels, which allows for overlapping execution of our reader~(\textit{read\_input}), executor~(\textit{compute\_mm}) and writer~(\textit{write\_result}) processes.

The HBM (High Bandwidth Memory) global memory of the FPGA holds any data incoming/outgoing from/to the host application. It provides independent AXI channels for access and communication between the Vitis kernels and the HBM PCs (pseudo channels) through a segmented crossbar switch network. Each PC represents a portion of the total available global memory and thus the entire HBM subsystem can facilitate high bandwidths because of the independent parallel access of each individual PC. The reader and writer processes written in HLS can interact with these AXI channels in order to further move data to or from the Montgomery Multiplier that resides inside the executor process as an RTL blackbox. 

This variant of the HLS wrapper is used by the majority of our Verilog RTL designs. For all three analyzed word sizes, 24/32/64 bits, we implemented the following types of designs that are plugged-in the HLS wrapper described above:

\begin{itemize}
  \item Row-Parallel design: the inner loops of the CIOS algorithm are unrolled and all partial products are computed through parallel DSP-based units, with a final DSP-based 384-bit adder for carry propagation; 
  \item Row-Serial design: the inner loops of the CIOS algorithm use a single pipelined DSP-based unit for computing the multiply-and-add operation through the sequential scheduling of all words of the operands; 
  \item Karatsuba-based design: this design evaluates the automatic usage of DSPs by the Vivado synthesizer through the use of the \textit{use\_dsp} flag; in order to optimize the multiplier usage we approach the Karatsuba integer multiplication instead of the classic schoolbook multiplication used in the previous designs - this version is evaluated only for the 32 and 64-bit sizes;  
  \item Full HLS design: we also evaluate the implementation of the CIOS algorithm fully in HLS, and control the usage of DSPs through the compiler's \textit{pragmas}. 
\end{itemize}

\begin{figure}[!htp]
\centering
    \includegraphics[width=1\textwidth]{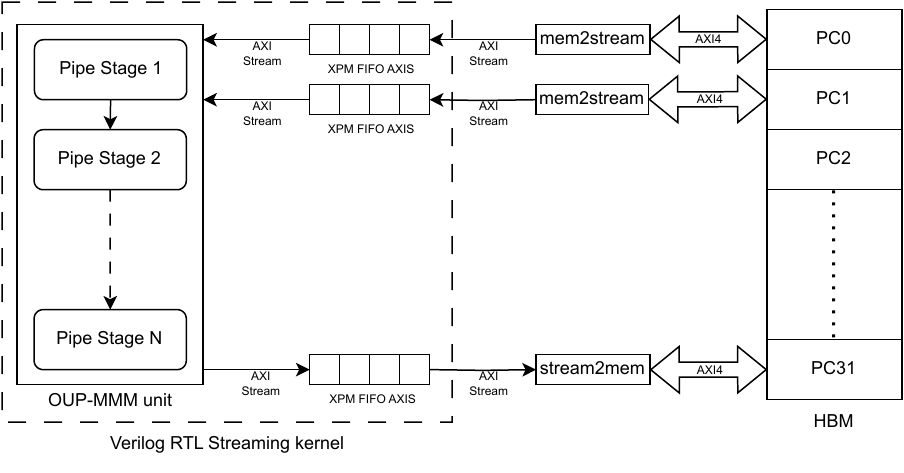}
    \caption{Second variant of the HLS wrapper, composed of the reading (mem2stream) and writing (stream2mem) tasks implemented as independent HLS kernels and the Verilog RTL kernel containing the OUP MMM unit with AXI Stream interfaces connected to the corresponding producers and consumer}
    \label{fig:hls_wrapper2}
\end{figure}

The second variant of the HLS wrapper, shown in Figure~\ref{fig:hls_wrapper2} differs in the fact that the read, compute and write functions are implemented as independent kernel objects instead of part of the same kernel. The motivation behind is the last design type analyzed, the Outer Unrolled Pipeline (OUP). This design cannot run in pipeline if used in the first variant of the wrapper, as the RTL backbox feature in Vitis 2023 is currently limited to be used in non-pipelined regions~\cite{rtl_blackbox}. This is not a problem for the design types enumerated earlier, as those ones are sequential designs and cannot process a new input until the current one is not finished, but for the pipelined design, using the first variant of the wrapper will cancel the benefits of processing multiple inputs at different stages.

Figures~\ref{fig:hls_wrapper1} and \ref{fig:hls_wrapper2} show an overview of the two wrapper designs used for accessing the global memory of the FPGA. The HBM memory in the Alveo U55C platform is divided in 32 pseudo-channels, each channel providing access to 512MBs of the total of 16GBs global memory. Kernel interfaces can be bounded to separate pseudo-channels in order to provide parallel access and better bandwidth usage. In out implementations, all three data interfaces (the two operands and the result) have their own pseudo-channel assigned. For the first variant of the HLS wrapper illustrated in Figure~\ref{fig:hls_wrapper1} we show the unified design where the \textit{read\_input} and the \textit{write\_result} tasks communicate with the global memory over the AXI4 protocol~\cite{axi_protocol}. These tasks further communicate with the \textit{compute\_mm} over HLS streams~\cite{hls_stream} channels. The \textit{compute\_mm} HLS task has an auto-generated control FSM (Finite State Machine) unit which commands the Verilog RTL blackbox that represents our Montgomery Multiplier unit. For each pair of operands read from the stream, we parallel load the 384-bits directly to the blackbox, and in the same manner, we parallel read the 384-bits result and push it to the output HLS stream queue. For communication with the RTL blackbox, the auto-generated control FSM of the \textit{compute\_mm} HLS task and the Verilog Montgomery Multiplier designs use the \textit{ap\_ctrl\_chain}~\cite{ap_ctrl_chain} protocol for the handshake mechanism.

Figure~\ref{fig:hls_wrapper2} shows the second variant used for the pipeline execution flow of the OUP-MMM unit. The \textit{read\_input} tasks are now independent kernels called \textit{mem2stream}, and the \textit{write\_result} task is replaced by \textit{stream2mem}. These kernels use the same HLS code as the read/write functions in the first variant. The Verilog RTL Streaming Kernel is written completely in Verilog and it incorporates the OUP-MMM unit and three XPM (Xilinx Parameterized Macros) AXI Stream (AXIS) FIFO buffers~\cite{xpm}. For the second variant, the communication between the Verilog RTL Streaming kernel and the read/write HLS kernels is done through AXI Stream interfaces instead of the HLS Stream ones used in the previous design. Similar to the first variant, we read each operand pair over two parallel AXI Streams and further parallel load the entire 384-bit values to the OUP-MMM unit. Similarly, for the result, we parallel read the entire 384-bit result and push it over AXI Stream to the output XPM FIFO queue. 

Table~\ref{tab:hls_wrapper_overhead} shows the overhead (LUTs, FFs and BRAMs) that the HLS wrappers and Verilog RTL AXI Stream interfaces and streaming queues will add over the analyzed MM units. The \textit{Read} column shows the total resource consumption of the two \textit{read\_input} HLS tasks for Variant I and the total resource consumption of the two \textit{mem2stream} kernels for Variant II. Similarly, The \textit{Write} column shows the resource consumption for the \textit{write\_result} HLS tasks for Variant II, and the resource consumption of the \textit{stream2mem} kernel for Variant II. For both variants, we have defined internal streaming queues (HLS and AXI Stream correspondingly) of depth 8 for each interface and an entry size of 512 bits. This results in a total of $3 \times 8 \times 512 = 1.5KB$ memory usage for all queues involved in the final design in both variants.

For the second variant, we have additional memory usage (implemented as \textit{LUT as Memory} for higher performance) for the three XPM FIFO AXIS queues that we instantiated inside the Verilog RTL Streaming kernel with a depth of 32 entries of 512 bits each (we observed that this depth allowed for continuous pipeline execution without causing a bottleneck). The \textit{Compute} column includes this resource consumption together with the logic needed for implementing the three AXI Stream interfaces inside the Verilog RTL Streaming kernel and is depicted as overhead for the second variant. For Variant I, the same column shows the overhead added by the \textit{compute\_mm} HLS task. For both variants, the tools estimate a maximum achievable operating frequency of 890MHz which means that the wrappers do not represent a frequency bottleneck for the entire multiplier design. All values are retrieved from post-placement and route implementation of the designs, overhead values being the same regardless of the multiplier design type used (Row-Parallel, Row-Serial, Outer Unrolled Pipeline) or the word size (24/32/64 bits).

\begin{table}[!htp]
    
    \centering
    \begin{tabular}{c|c|c|c|c}
       HLS wrapper & Read  & Compute  & Write & Total \\\hline
       Variant I  & \makecell{2857 CLB LUTs \\ 3871 CLB Regs \\ 46 BRAMs} & \makecell{1051 CLB LUTs \\ 2607 CLB Regs \\ 0 BRAMs}  & \makecell{2620 CLB LUTs \\ 2788 CLB Regs \\ 8 BRAMs}  & \makecell{6528 CLB LUTs \\ 9266 CLB Regs \\ 54 BRAMs} \\\hline
       Variant II  & \makecell{3889 CLB LUTs \\ 5846 CLB Regs \\ 13 BRAMs} & \makecell{851 CLB LUTs \\ 2758 CLB Regs \\ 0 BRAMs} & \makecell{2160 CLB LUTs \\ 3419 CLB Regs \\ 8 BRAMs} & \makecell{6900 CLB LUTs \\ 12023 CLB Regs \\ 21 BRAMs} \\
    \end{tabular}
    \caption{HLS wrapper overhead}
    \label{tab:hls_wrapper_overhead}
\end{table}

In the next two sections, we further describe the internal workings of each type of Montgomery Multiplier and DSP-only arithmetic units implemented, together with the throughput results obtained, the resource consumption, power and frequency achieved. 

\section{Hardware design implementations}
\label{sec:design}

We divide this section into three parts for a better understanding of the implemented Montgomery multipliers. The first part goes and details all the DSP-based arithmetic units used for the word integer operations: multiply, add, multiply-and-add, multiply-and-add-and-carry. Besides using DSPs, the word arithmetic units also make use of FDRE primitives~(single D-type flip-flops) and/or SRL16E primitives~(shift register look-up tables) to create delay lines for pipelining the inputs and internal results. All arithmetic units are implemented for each of the evaluated word sizes: 24/32/64-bits.

In the second part we further detail the design structure of the CIOS-based Verilog-implemented Montgomery multipliers, going through the three main design types: Row-Parallel, Row-Serial, Outer Unrolled Pipeline, and detail how each of these design types use the DSP-based arithmetic units described in the first part.

The third part details the designs that rely on tools: the complete HLS implementation, and the naive Verilog implementation that uses the synthesizer's flag for automatic inference of DSPs. For the later design approach, we use for the implementation of the word multiplications a Karatsuba multiplier that operates on either 64 or 32-bits and decomposes the operands down to 16 bits in order to fit in a DSP. The Karatsuba multiplier is implemented using the addition, subtraction and multiplication operators from Verilog.

\subsection{Word arithmetic units}

The main operations involved in the CIOS algorithm shown in Algorithm~\ref{cios_mmm} are the addition and multiplication of the words composing our large integer operands and modulus. The most demanding step implies the following operation: $(carry, t[i+j]) = a[j] \times b[i] + t[i+j] + carry$, as it involves a multiplication, a self-add and an extra addition with the carry, resulted from a previous operation. Apart from that we also have a simple word self-addition involved in the operation $t[i+s] = t[i+s] + carry$ and a simple word multiplication in the operation $m = t[i] \times p'$. A naive approach would mean implementing either one or two-word multipliers depending on the level of parallelism wanted, and at least one-word adder, or up to three if pipeline is desired, to cover all word operations enumerated earlier.   

As we aim to efficiently use the available DSPs from a resource point of view we propose the following arithmetic units based on DSPs, for executing all the needed operations from the CIOS algorithm that are employed in our Montgomery multiplier designs (we denote $X$ as being the size of the word which can be 24/32/64-bits) :

\begin{itemize}
  \item MUL\_$X$: a pipelined multiplier that performs $P = A\times B$ where $A,B$ are $X$-bits words and $P$ is a $2X$-bit word 
  \item ADD384: an adder that performs $P = A + B$ where $P,A,B$ are 384-bits operands  
  \item MADD\_$X$: a pipelined multiply-and-add unit that performs $P = A\times B + C$ where $A,B,C$ are $X$-bits words and $P$ is a $2X$-bit word
  \item MADD384\_$X$: a multiply-and-add unit that performs $P = A\times B + C$ where $A$ is a 384-bit input, $B$ is a $X$-bit word and $P,C$ are $(384+X)$-bit values
  \item MADDCARRY\_$X$: a multiply-and-add-and-carry unit that performs $(CARRY, P) = A\times B + C + CARRY$ where $P,A,B,C$ are $X$-bits words
\end{itemize}

\begin{figure}[!htp]
\centering
    \includegraphics[width=0.9\textwidth]{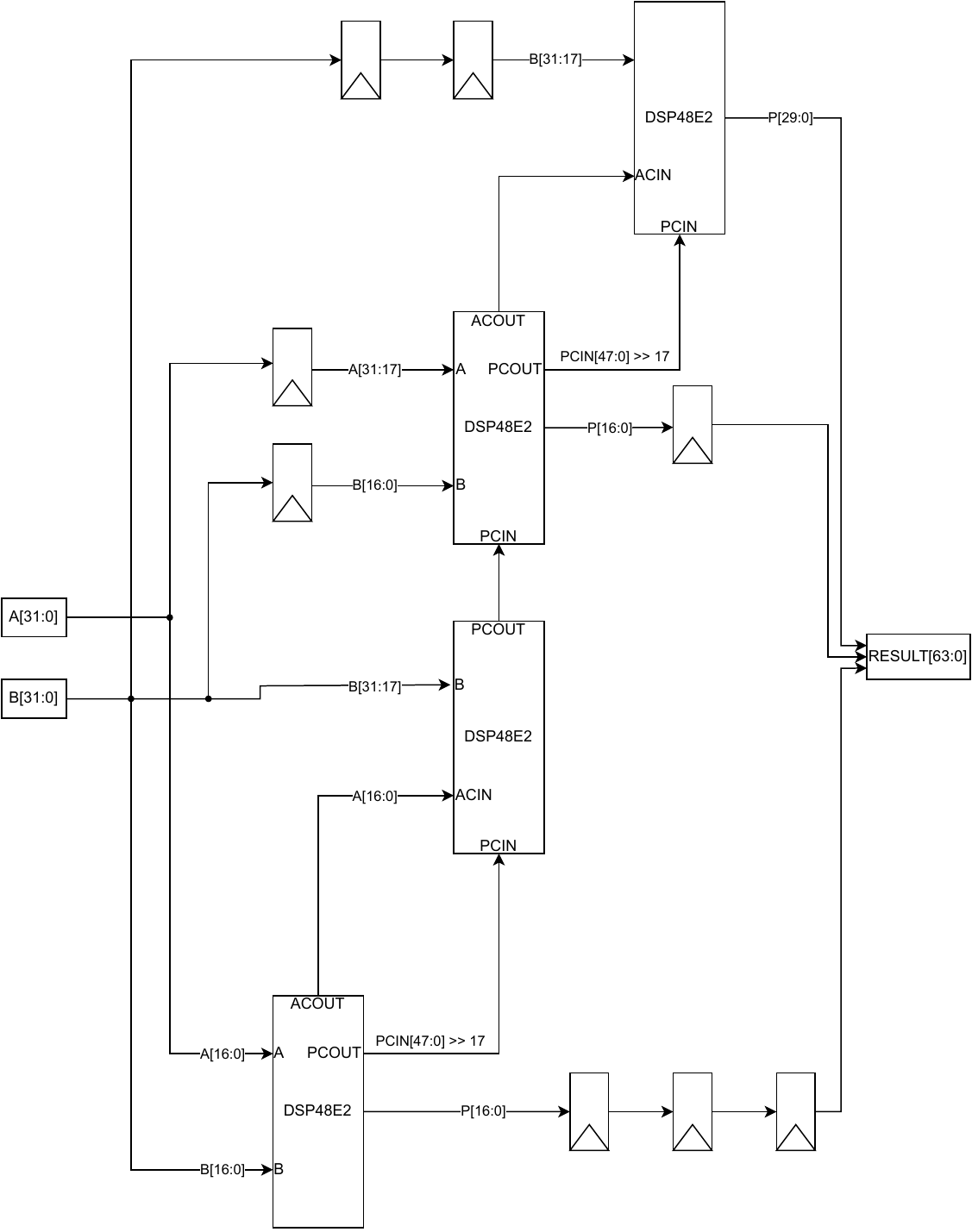}
    \caption{MUL\_32 arithmetic unit using four DSPs, and additional registers (FDRE) for pipeline implementation}
    \label{fig:dsp32_mul}
\end{figure}

We begin with the MUL\_$X$ unit which is based on the classic schoolbook multiplication. As the DSP48E2 unit uses a signed rectangular multiplier of $27\times 18$-bits, and also comes with the ability of right-shifting by 17 bits a cascaded output from a previous DSP unit in the same column, we split our $X$-bit inputs into 17-bit words in order to efficiently use and cascade the DSP units to perform the desired multiplication of our $X$-bit words. By the use of the internal input DSP registers A1, A2, B1, B2 which can be cascaded as well, and also through FDRE and SRL16E-based delay lines where needed, we synchronize our split inputs to achieve a pipeline design that can output at each clock cycle a new output. We also use both M and P internal registers of the DSP to achieve the maximum possible frequency of the DSP unit even if this comes with the penalty of an increased latency. Figure~\ref{fig:dsp32_mul} shows a 32-bit pipelined multiplier using 4 DSPs, and additional CLB registers to pipeline the inputs and outputs. The internal DSP registers A1, A2, B1, B2 are also enabled to properly pipeline and synchronize the inputs but are omitted here. A similar approach is done for the 24 and 64-bit word cases.   

The ADD384 unit is made of 8 parallel DSPs as we can efficiently make use of the 48-bit adder, cascaded through the CARRYCASCIN and CARRYCASCOUT signals of each two neighbor primitives. The unit does not operate in pipeline as the carry signals and the output values are not registered. The reason behind this design choice was to have all 8 DSPs in a single column cascaded through the mentioned signal pair. As the multiplier is not needed, the DSPs need two stages, one to register the inputs, and one to register the output. The final latency of the entire adder is 9 cycles, two cycles to perform all parallel additions, and another 7 cycles for the carry bits to propagate accordingly, as each carry is directly connected to the 48-bit adder. It requires an additional cycle on each DSP to update the final result.  

The MADD\_$X$ unit enhances MUL\_$X$ to implement the multiply-and-add operation through the use of the 48-bit C port of the DSP. For the 24 and 32-bit variants, this is straightforward, we assign the third operand to the C port of the first DSP in the chain which computes the least significant bits of the final result. Both these sizes fit in the 48-bit C port and do not overflow the 48-bit PCOUT port which is further used in the DSP chain. The reason behind this comes from the fact that, in both MUL\_24 and MUL\_32 variants, the first DSP computes a $17\times 17$-bit multiplication which results in 34-bits. Even with the addition of a 24 or 32-bit value coming from the third operand of the multiply-and-add operation, the output of this DSP does not exceed 35-bits, which can be further cascaded and right-shifted into the second DSP without affecting the final result. 

For the 64-bit variant, we have an extra DSP added to the MUL\_64 unit as the addition of the 64-bit operand is done through two DSPs, while also maintaining the pipeline behavior. The least significant 17 bits of the operand are fed into the first DSP in the chain which further outputs the least significant 17-bits of the final result. After that, the full 35-bit result is cascaded and right-shifted into the second DSP. The second DSP performs its normal operation and cascades its result further into the third DSP which behaves the same as in the normal 64-bit multiplication implementation. 

In the simple 64-bit multiplication implementation based on DSPs, the least significant 17 bits of the result of the third DSP are forwarded to the final output port on bit positions 17 to 33. In the case of the MADD\_64 unit, we forward the entire result of the third DSP into the extra DSP in order to add the remaining most significant 47-bits of our third operand. The result of the extra DSP will not exceed 48-bits as we have an addition between a 47-bit value and a 35-bit value, and this result is further cascaded to the following DSP which continues the normal in-chain operation as in the initial MUL\_64 unit without causing any carry loss in the final result. Equation~\ref{MADD64} shows the execution chain of the first five DSPs of the MADD\_64 unit, where $A,B,C$ are the three 64-bit operands of the multiply-and-add operation, $P0-4$ are the output ports of the five DSPs, and $RESULT$ is the final ouptut of the MADD\_64 unit.   

\begin{equation} \label{MADD64}
\begin{gathered}
DSP[0] \rightarrow P0[34:0] = A[16:0] \times B[16:0] + C[16:0] \rightarrow RESULT[16:0] = P0[16:0]\\  
DSP[1] \rightarrow P1[34:0] = A[16:0] \times B[33:17] + P0[34:0] >> 17\\
DSP[2] \rightarrow P2[35:0] = A[33:17] \times B[16:0] + P1[34:0]\\
DSP[3] \rightarrow P3[47:0] = P2[35:0] + C[63:17] \rightarrow RESULT[33:17] = P3[16:0] \\
DSP[4] \rightarrow P4[34:0] = A[50:34] \times B[16:0] + P3[47:0]>>17 \\
\end{gathered} 
\end{equation}

The MADD384\_$X$ unit is built upon the MADD\_$X$ and ADD384 units described earlier in order to provide a solution for unrolling the inner loop of the CIOS algorithm. As a reminder, the inner loop of the CIOS algorithm is shown in Algorithm~\ref{cios_inner_loop} and our MADD384\_$X$ unit computes the following operation: $P = A\times B + C$ where $A$ is a 384-bit input, $B$ is a $X$-bit word and $P, C$ are $(384+X)$-bit values. The extra $X$-bits from $P$ and $C$ represent the position $i+s$ in our $t$ array used in the algorithm for storing the partial products.

\begin{algorithm}[!htp]
\caption{CIOS inner loop}\label{cios_inner_loop}
\begin{algorithmic}
\State Let $s$ be the number of $X$-bit words we split our 384-bit operands into
\State Let $T$ be an array of size 2$s$ $X$-bit words
\State Let $CARRY = 0$ be a $X$-bit word
\State Let $i$ be the current iteration of the outer loop
\For{\texttt{$j \gets 0$ to $s$}}
    \State $(CARRY, T[i+j]) = A[j] \times B[i] + T[i+j] + CARRY$
\EndFor
\State $T[i+s] = T[i+s] + CARRY$ \Comment{Propagate extra $carry$ to next word}
\end{algorithmic}
\end{algorithm}

The unit uses $s$ parallel MADD\_$X$ units in order to compute $temp[j] = a[j] \times b[i] + t[i+j]$, where $temp[j]$ is a $2X$-bit word. In other words, the first stage of the unit computes the multiplication between an entire 384-bit operand by the current scanned $i$-th word of the second operand and adds to this result our previous partial products stored in the $t$ array. All the carries of our products sit in the higher half of each $temp[j]$ result. Before saving the new results into the $t$ array, we send all $temp[j]$ results with $j= 0,1,\cdots,s-1$ to the ADD384 unit, where the temporary results are added to themselves shifted by $X$ bits. This will add the higher half of $temp[j]$ to the lower half of $temp[j+1]$, taking care thus of the carry propagation needed in the inner loop of CIOS.

\begin{figure}[!htp]
\centering
    \includegraphics[width=1\textwidth]{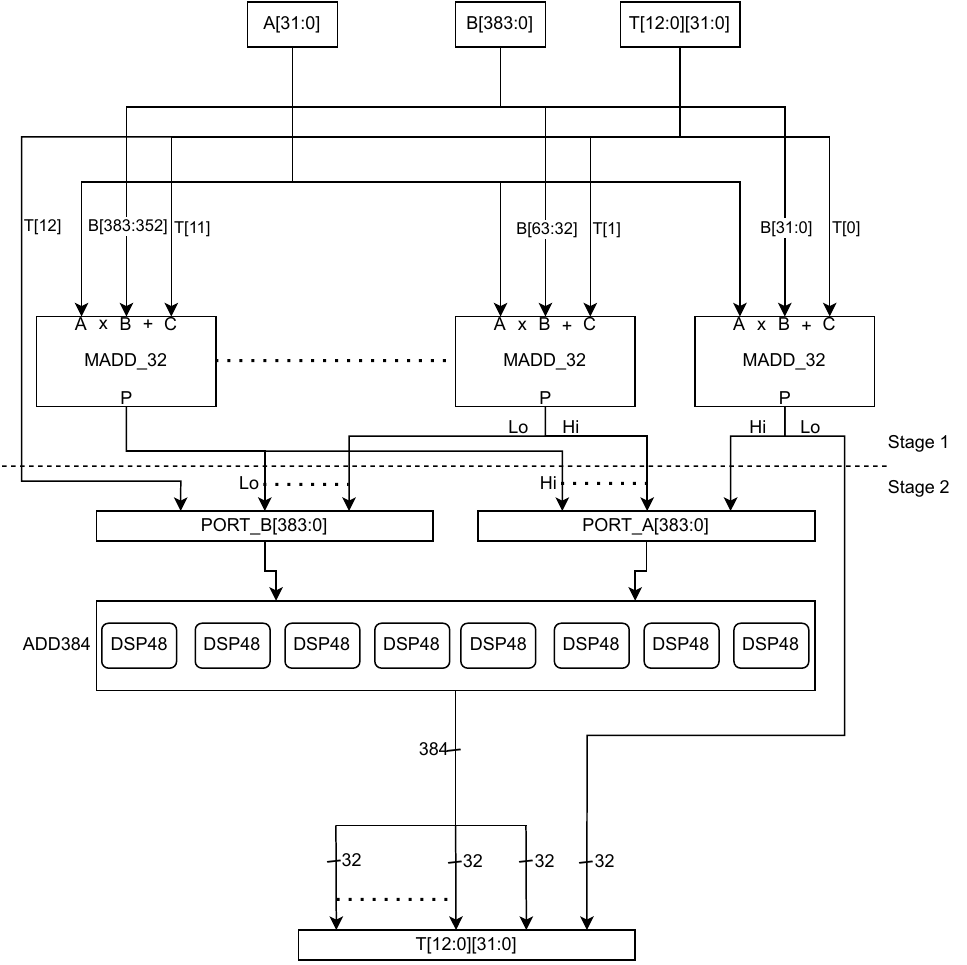}
    \caption{MADD384\_32 arithmetic unit using 12 parallel MADD\_32 units and an ADD384 unit. The T array is implemented as $2s$ LUTRAM 32-bit registers in order to have parallel read and parallel write access to all registers at once to avoid extra latency.}
    \label{fig:madd384_32}
\end{figure}

As we also need to take care of the final carry addition that happens after the inner loop which computes $t[i+s] = t[i+s] + carry$, we assign the two inputs of the ADD384 unit as in Figure~\ref{fig:madd384_32}. The figure shows for the 32-bit variant the parallel load of the first word scanned of operand A, the entire operand B, and the first $s+1$ (13 in our case) words of the $T$ array that holds the partial products. After we obtain the temporary results of each MADD\_32 unit, we assign all $Hi$ halves (32-bits) to PORT\_A of the ADD384 unit, and all $Lo$ halves (32-bits) to PORT\_B. The lower half of the first MADD\_32 unit is forwarded directly to $T[0]$ from the array, as $CARRY$ will always be zero before execution of the inner loop. To fill the remaining 32-bits of PORT\_B, we shift all lower halves by a position to the right and fill the most significant 32 bits of PORT\_B with the extra word of the $T$ array. With this alignment, we take care of the step $t[i+s] = t[i+s] + carry$ which happens immediately after the inner loop. With these two stages, the multiply-and-add and the 384-bit addition, we compute Algorithm~\ref{cios_inner_loop} within a single MADD384\_$X$ unit. As ADD384 is not pipelined, MADD384\_X is also not operating in the pipeline, but one could add pipeline registers between the two stages and inside the ADD384 unit. We decided to not go with this approach in order to save resources and obtain better placement and frequency.

\begin{figure}[!htp]
\centering
    \includegraphics[width=1\textwidth]{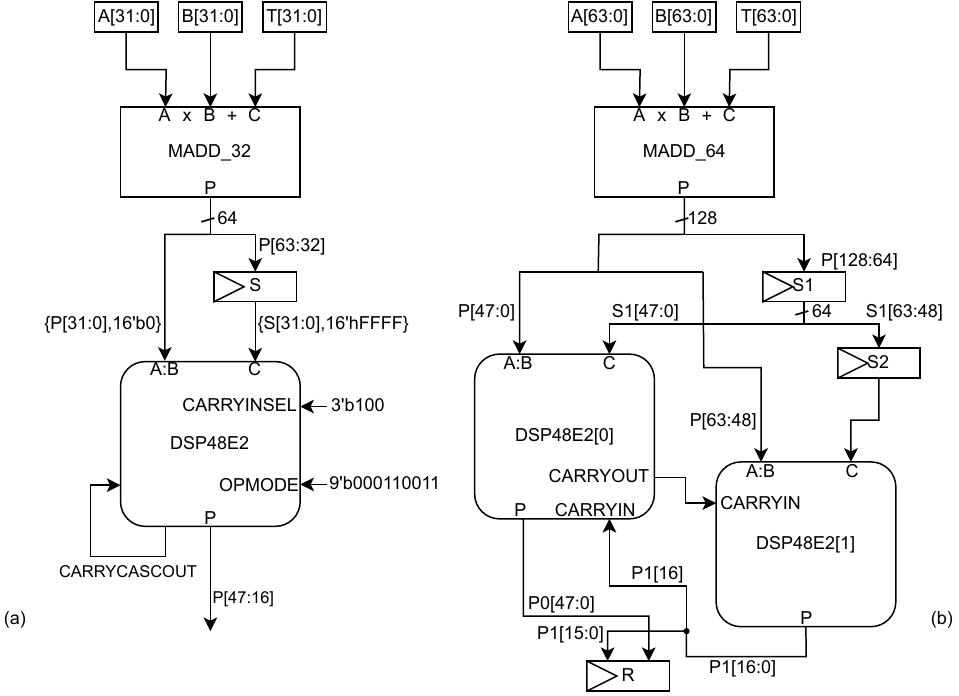}
    \caption{MADDCARRY\_32 arithmetic unit (a) and MADDCARRY\_64 arithmetic unit (b)}
    \label{fig:maddcarry_32_64}
\end{figure}

Our last unit, MADDCARRY\_$X$, enhances the MADD\_$X$ unit by storing the carry resulting from the multiply-and-add operation and using it for the next pair of inputs. This design uses an extra DSP for the 24/32-bit variants, and two extra DSPs for the 64-bit operands, but allows for a pipeline scheduling of all words scanned throughout the execution of a CIOS inner loop as we can send a new pair of inputs each clock cycle. Figure~\ref{fig:maddcarry_32_64} shows how the extra DSPs are used for the 32 and 64-bit variants together with extra CLB registers in order to accomplish the execution of the operation $(CARRY, T) = A\times B + T + CARRY$ encountered in the inner loop. We save the higher half(the carry of the multiply-and-add operation) in a register that delays it by one cycle in order to have it available for the next result. This design allows us to use only one or two additional DSP units to obtain the results of an inner loop execution with a similar latency as the MADD384\_$X$ unit, but using only one MADD\_$X$ unit and no ADD384 unit.

Table~\ref{tab:word_unit_overhead} shows the number of DSP primitives and CLB Registers used by each implemented word arithmetic unit based on the bit-size variant. We also show the latency of each unit in clock cycles and whether the specific unit can run in pipeline or not. Even though the MADD384\_$X$ is based on the MADD\_$X$ unit which is pipelined, as described in this section, this unit also uses in its second stage the ADD384 unit, which is not pipelined in order to save resources. The Row-Parallel Montgomery multiplier which uses this kind of word arithmetic unit is also not designed to process new inputs in a pipeline fashion. The resources are retrieved from the post-synthesis stage, the \textit{CLB Regs} column refers to FPGA 1-bit D-type flip-flops implemented through FDRE primitives and the \textit{Pipelined} column refers to the fact that each design can output a new result every clock cycle. The \textit{LUTs} column is important for the 64-bit implementation, as this uses SRL16E primitives for creating the delay lines instead of FDREs, mainly because we need to delay values up to 17 cycles, and using SRL16Es offers better routing and placement for the synthesis and implementation tool compared to using chained FDREs.  

\begin{table}[!htp]
    \centering
    \begin{tabular}{c|c|c|c|c|c|c}
       Word arithmetic unit & Bit Size & DSPs & CLB LUTs & CLB Regs & Latency & Pipelined \\\hline
       MUL\_$X$ & \makecell{24-bit \\ 32-bit \\ 64-bit}  & \makecell{2 \\ 4 \\ 16} & \makecell{0 \\ 0 \\ 422} & \makecell{17 \\ 130 \\ 473} &  \makecell{4 \\ 6 \\ 18} & YES \\\hline
       MADD\_$X$ & \makecell{24-bit \\ 32-bit \\ 64-bit}  & \makecell{2 \\ 4 \\ 17} & \makecell{0 \\ 0 \\ 486} & \makecell{41 \\ 162 \\ 554} &  \makecell{4 \\ 6 \\ 20} & YES \\\hline
       MADD384\_$X$ & \makecell{24-bit \\ 32-bit \\ 64-bit}  &  \makecell{40 \\ 56 \\ 110} & \makecell{0 \\ 0 \\ 2916} & \makecell{656 \\ 1944 \\ 3324} &  \makecell{13 \\ 15 \\ 29} & NO \\\hline
       MADDCARRY\_$X$ & \makecell{24-bit \\ 32-bit \\ 64-bit}  & \makecell{3 \\ 5 \\ 19} & \makecell{0 \\ 0 \\ 486} &  \makecell{67 \\ 196 \\ 700} &  \makecell{6 \\ 8 \\ 23} & YES \\\hline
        ADD384 & 384 & 8 & 0 & 0 & 9 & NO \\\hline
    \end{tabular}
    \caption{Number of DSPs, LUTs, Registers (FFs), and latency (clock cycles) for each word arithmetic unit based on supported bit size (post-synthesis)}
    \label{tab:word_unit_overhead}
\end{table}

\subsection{DSP optimized Montgomery multipliers}
\label{main_mmm_sec}

\begin{figure}[!htp]
\centering
    \includegraphics[width=1\textwidth]{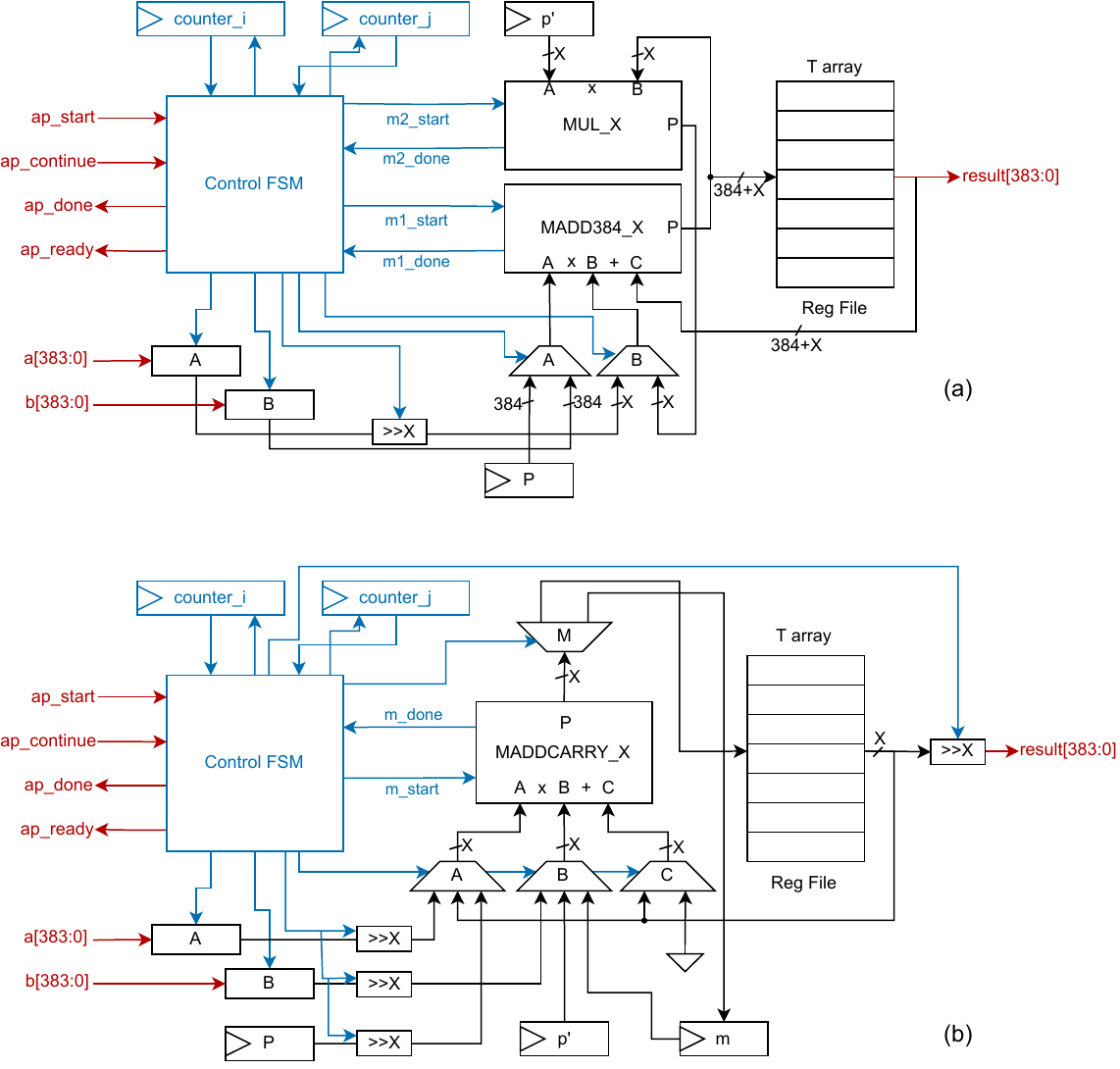}
    \caption{Row-Parallel (a) and Row-Serial (b) design for Montgomery multiplier. Blue represents the control path, red represents the input/output paths, and black represents the internal data path. $X$ represents the size of the words employed by the three size variants implemented: 24/32/64-bits}
    \label{fig:mont_mult384}
\end{figure}

The Verilog Montgomery multipliers where we manually instantiate and optimize the DSP unit usage are called: Row-Parallel (RP), Row-Serial (RS), and Outer Unrolled Pipeline (OUP). Figure~\ref{fig:mont_mult384} shows the implementation overview for the Row-Parallel and Row-Serial designs. The Control FSM unit is similar for both variants, wherein the initial state it waits for the \textit{ap\_start} signal to be asserted while keeping the \textit{ap\_ready} logic positive to flag an upstream module that it is idle and it can process a new input. Once the upstream module asserts the \textit{ap\_start} signal, the control unit parallel loads the two operands, and resets all counters and internal registers in order to start the CIOS algorithm. These signals are part of the \textit{ap\_ctrl\_chain} handshake used to communicate with the parent HLS task through the RTL blackbox feature. 

Based on the values of \textit{counter\_i} and \textit{counter\_j}, it tracks the current iteration for the outer loop and the inner loop. As we have two inner loops, the control unit moves from the initial state to a state which we call $LOOP\_1$ that processes the first inner loop, afterwards, it moves further to a third state similarly called $LOOP\_2$ that takes care of the second inner loop. After complete execution of the second inner loop, we jump back to state $LOOP\_1$ where we first verify the counter corresponding to the outer loop, and check if we reached the final iteration. If that is the case we jump to the final state where we output the final results stored in the $T$ array of registers and assert the \textit{ap\_done} signal, to flag a downstream module that we have an output ready to be processed. Otherwise, if we don't reach the final iteration of the outer loop we go through the $LOOP\_1$ and $LOOP\_2$ again. Both $LOOP\_1$ and $LOOP\_2$ states are complete when the inner loop counter that is shared between them marks the final iteration.

For the Row-Parallel design, we use a MADD384\_$X$ unit a simple MUL\_$X$ unit. The name is inspired by the fact that through the MADD384\_$X$ unit, we can compute in parallel a row of partial products as we described the use of this unit in the previous subsection. The additional multiplier is needed for computing the quotient used in the second inner loop for performing the reduction phase. This design has a separate state in the control unit between $LOOP\_1$ and $LOOP\_2$ where the control unit starts the simple multiplier and waits until the result is produced before jumping to the second inner loop state. 

As the MADD384\_$X$ contains a large number of DSPs, it is shared between the two inner loops for resource saving, and controlled, like the simple multiplier, through the start and done signals shown in Figure~\ref{fig:mont_mult384}. The unit parallel loads an entire 384-bit operand, and $384+X$-bits from the $T$ array, which is why the $T$ array is implemented as a LUTRAM register file of $2s$ parallel $X$-bit registers in order to parallel read and write from and to them in a single cycle. After the completion of the first inner loop, the least significant word (as required by the CIOS algorithm) computed by the MADD384\_$X$ unit is forwarded to the MUL\_$X$ unit so that the quotient calculation can start in the very next clock cycle. We have both the modulus $P$ and the prime modulus $p'$ pre-computed and stored in ROM-style registers.

The Row-Serial design emphasizes on the MADDCARRY\_$X$ unit and uses only one for both the inner loop and also for the quotient calculation, making it a compact design while also keeping a close latency to the Row-Parallel. Instead of computing all partial products of an inner loop in parallel, we compute them in a pipeline fashion, obtaining a new partial product each clock cycle. Thanks to the MADDCARRY\_$X$ unit design which stores the carry of the previous iteration internally as described in the previous section, it allows us to use it immediately in the next clock cycle for the current iteration without any additional delay.

\begin{figure}[!htp]
\centering
    \includegraphics[width=1\textwidth]{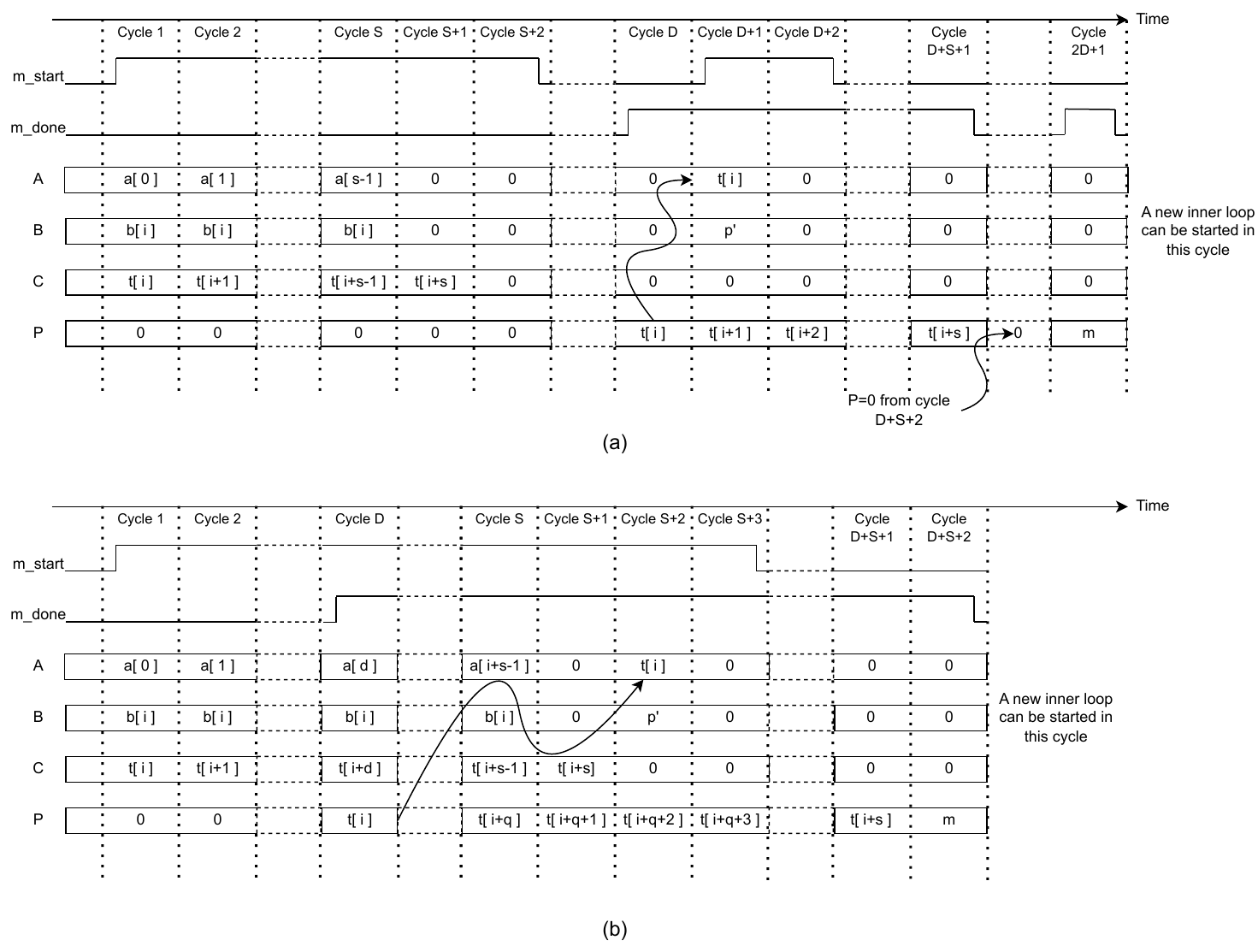}
    \caption{Pipeline scheduling of a inner loop for iteration $i$ of the outer loop for MADDCARRY\_64 (a), MADDCARRY\_24 and MADDCARRY\_32 (b)}
    \label{fig:maddcarry_scheduling}
\end{figure}

Figure~\ref{fig:maddcarry_scheduling} shows the word scheduling implemented in the control unit, where all pair of words $(a[j], b[i], t[i+j])$ assigned to ports $A, B, C$ of the MADDCARRY\_$X$ unit, with $i$ fixed and $j=0,1,\cdots,s-1$, are sent one by one, within a clock cycle from each other for the execution of a inner loop. The port $P$ in the figure shows at each clock cycle where the result of the MADDCARRY\_$X$ unit will be forwarded: either to the $t$ array of registers or the register $m$ for the quotient. With \textit{m\_start} and \textit{m\_done} we define the control signal that starts the unit and the status signal that flags when a result is ready to be read.

The figure shows the execution for all three size variants, with the 24 and 32-bit variants behaving the same, while the 64-bit variant has a separate scheduling. With $D$ we define the latency of the MADDCARRY unit for outputting the first result. The different behavior comes from the relation between the value $D$ and the value $S$ (the number of words). The relation between these two values affects the moment we have our first partial product $t[i]$ computed, and ready to be used for calculating the quotient $m$ . 

For the 24 and 32-bit variants, $S>D$, meaning that we get our first partial product before finishing sending all input pairs to the unit. Once, all pairs are being sent, we can immediately send $t[i]$ as in Figure~\ref{fig:maddcarry_scheduling}(b), together with the prime modulus $p'$ (cycle $S+2$), as the input ports are available to be used (we keep the $m\_start$ signal asserted). Afterward, we push in the next cycle, $S+3$, with $m\_start$ still being asserted a pair of logic zeroes on the input ports in order to reset the internal state of the unit and have it ready for a following inner loop. Pushing '0' at the input ports of the MADDCARRY\_$X$ unit will push at the output the value present in the registers used for holding the current carry while also resetting these registers at the same time. Cycle $S+1$ shows how the final carry addition that needs to be added to word $t[i+s]$ can be performed by pushing zeroes on the ports $A,B$ and the current $t[i+s]$ on port $C$. From the last input pair being sent for computing the quotient, we wait for $D$ cycles to get $m$, by that time, all $t[i]$, with $i=0,1,\cdots,s$ are also computed.

 For the 64-bit variant on the other hand, $D>S$, meaning that we get to send all our input pairs first, and afterward the MADDCARRY unit will output the first $t[i]$ after a delay equal to $D-S$ cycles. Because of this, once we push the set of inputs for computing $t[i+s]$ in cycle $S+1$, we push zeroes to the input ports in order to clear the accumulated carry in advance for when we are ready to compute $m$. Once we have the first $t[i]$ computed in cycle $D$, we use it in the next cycle to start the computation of $m$ without any further delay. All $t[i]$ with $i=0,1,\cdots,s$ are computed by cycle $D+S+1$, but as the computation of $m$ was started after $D$ cycles, we have to wait for that amount of time until cycle $2D+1$ to also get the $m$ computed. As we send out a pair of logic zeroes in cycle $D+2$, our unit is also cleared and ready to be used in the next cycle $2D+2$. Without clearing the unit, any new input that we would send to it will use as carry the last value that is available in the carry registers. Pushing zeroes to all ports of the unit acts as a reset of the internal registers.  

Because we operate on word sizes of $X$-bits, we can use a simple dual-port memory implementation for the $T$ array in order to relax the congestion of the routing and possibly improve the critical paths and final achieved frequency. For the Row-Serial implementation, we evaluate two separate cases to see how the performance is impacted: an implementation uses the LUTRAM-style of registers for the memory implementation like in the case of Row-Parallel in order to provide a fast one-cycle read and write. As we can access any register at any time this allows us to parallel access all registers at once in the final state where we want to output all 384-bits of the result at once in a single cycle to improve latency. 

The second variant uses a BRAM implementation based on the Xilinx Parameterized Macros (XPM) configured with a two-clock cycle read latency to improve clock-to-out timing and achieve the maximum frequency supported by the BRAM block. The "$>>X$" units in Figure~\ref{fig:mont_mult384} are used to shift and output each $X$-bit word needed in the current iteration of an inner loop, based on the value of the inner loop counter. We also use such a unit in the final state for the BRAM case where we cannot parallel load all $s$ words at once for the final result like in the case of Row-Parallel or LUTRAM Row-Serial, so we use the right-shift by $X$ unit to load all $s$ words one by one in a parallel register and output it once all results are available. We evaluate if this BRAM approach can improve the frequency at the expense of the additional clock cycles.   

\begin{figure}[!htp]
\centering
    \includegraphics[width=1\textwidth]{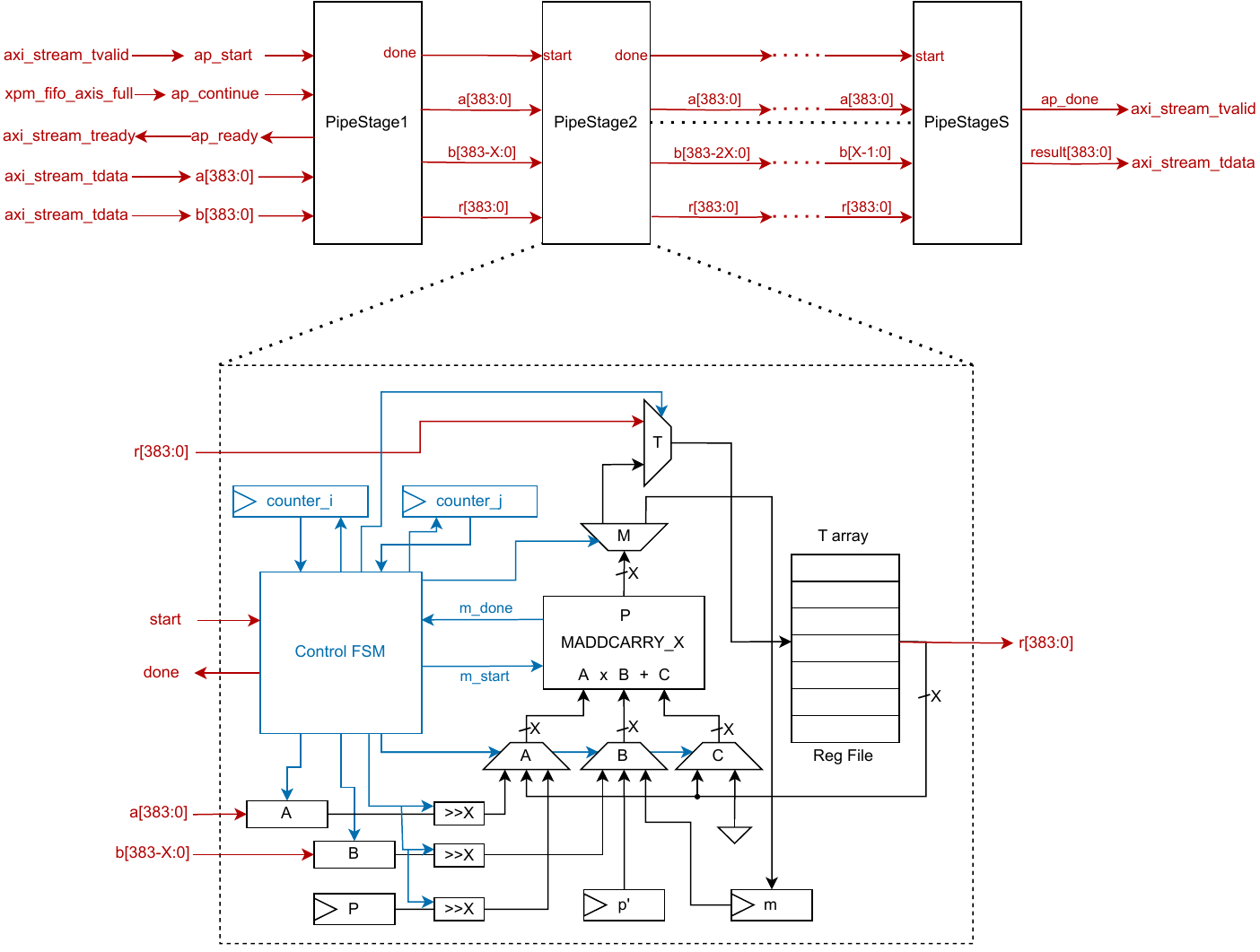}
    \caption{Outer Unrolled Pipeline design, composed of $s$ stages, showing all signal connections between each stage, signal connections to corresponding input and output AXI Stream FIFOs, and internal representation of a single pipeline stage}
    \label{fig:mont_mult384_pipe}
\end{figure}

Figure~\ref{fig:mont_mult384_pipe} shows the architecture design of the Outer Unrolled Pipeline (OUP) design which unrolls and offers pipeline execution of each iteration of the outer loop. The total number of stages is equal to $s$, the number of words the 384-bit operands are split into. Each stage transfers to the next one a full 384-bit operand, a right-shifted by $X$-bits operand, and the 384-bit intermediate result of the current iteration. The stages are controlled by the start/done pair of signals. The first stage has an \textit{ap\_ctrl\_chain} block-control interface used to start the first operation, and also stall the execution of the entire pipeline. The interface is forwarded to the AXI Stream FIFOs used for streaming the input operands into the pipeline. Similarly the last stage asserts the \textit{ap\_done} signal of the block-control interface, which is forwarded to the output AXI Stream FIFO. The pipeline stall is controlled by the full flag signal of the output AXI Stream FIFO identified by the name \textit{xpm\_fifo\_axis\_full}.

Each pipeline stage is implemented based on the LUTRAM-style Row-Serial design, being able to process both inner loops and the quotient calculation of an outer loop iteration. The pipeline stage is blocking, meaning that no other input can be processed until both inner loops are computed within the current stage. This design choice was made in order to preserve resource usage, with the penalty of having a latency equal to the processing of the two inner loops for each pipeline stage. As we use the LUTRAM variant, we can parallel load the entire 384-bit $r$ input into the $T$ array of $s$ registers, so that the restart of the pipeline stage for the next pair of inputs takes a single cycle, and the final state can also output the result to the next stage in a single cycle as well for a better latency. 

Table~\ref{tab:mmm_unit_overhead} shows the resource consumption and latency in clock cycles, post-synthesis (Flow\_PerfOptimized\_high strategy), of each main Montgomery Multiplier unit based on the bit-size variant. These results are retrieved for the core multipliers, without the overhead of the HLS/AXI Stream wrappers used in the final end-to-end designs which are depicted in Section~\ref{sec:results}. As described through Figures~\ref{fig:mont_mult384} and~\ref{fig:mont_mult384_pipe}, the core multiplier includes the word arithmetic unit(s) employed by the specific design, the FSM control unit for implementing the CIOS algorithm and interfacing through the \textit{ap\_ctrl\_chain} block-control protocol and the registers for storing the 384-bit input and output data. 

For the Row-Serial we also added the results of the BRAM-based implementation, which replaces the register file $T$ (the $t$ array of partial products in the CIOS algorithm) implemented with distributed LUTRAM, with a BRAM block. As we wanted to preserve the latency, we kept the local registers that store the two 384-bit input operands as distributed LUTRAM in order to have a fast one-cycle parallel load. For the Outer Unrolled Pipeline design, the \textit{Latency} column shows the number of clock cycles for both the first available result and in parentheses the latency for the next result in the pipeline. 

As we observe, the 64-bit version of all multipliers tends to have better latency, and this is expected, as even if the word arithmetic units involved for 64-bits have higher latency compared to the 24 and 32-bit versions (Table~\ref{tab:word_unit_overhead}), the number of CIOS iterations for the outer loop has a greater impact on the final design performance. Regarding resource consumption, the 24-bit version has better resource usage overall, except for the pipeline design, as again the 64-bit version has the least number of stages (6), dictated by the outer loop iterations, compared to the 24-bit (16 stages) and 32-bit (12 stages) versions. 

\begin{table}[!htp]
    \centering
    \begin{tabular}{c|c|c|c|c|c|c}
       Montogmery Multiplier & Bit Size & DSPs & CLB LUTs & CLB Regs & BRAM & Latency \\\hline
       
       Row-Serial & \makecell{24-bit \\ 32-bit \\ 64-bit}  & \makecell{3 \\ 5 \\ 19} &  \makecell{1050 \\ 1199 \\ 1724} &  \makecell{1748 \\ 1919 \\ 2585} & \makecell{0 \\ 0 \\ 0} & \makecell{802 \\ 554 \\ 494}  \\\hline
       
       Row-Serial (BRAM) & \makecell{24-bit \\ 32-bit \\ 64-bit}  & \makecell{3 \\ 5 \\ 19} &  \makecell{842 \\ 955 \\ 1636} &  \makecell{1383 \\ 1556 \\ 2216} & \makecell{1 \\ 1 \\ 1} & \makecell{917 \\ 641 \\ 557}  \\\hline
       
       Row-Parallel & \makecell{24-bit \\ 32-bit \\ 64-bit}  & \makecell{42 \\ 59 \\ 120} &  \makecell{1692 \\ 1622 \\ 4865} &  \makecell{2707 \\ 4116 \\ 5825} & \makecell{0 \\ 0 \\ 0} & \makecell{497 \\ 421 \\ 427}  \\\hline
       
       Outer Unrolled Pipeline & \makecell{24-bit \\ 32-bit \\ 64-bit}  & \makecell{48 \\ 60 \\ 114} &  \makecell{14050 \\ 10490 \\ 8753} &  \makecell{33679 \\ 27282 \\ 17401} & \makecell{0 \\ 0 \\ 0} & \makecell{844(52) \\ 576(48) \\ 504(84)} \\\hline
    \end{tabular}
    \caption{Resource consumption and latency (clock cycles) for each bit-size variant of the main Montgomery multiplier units (post-synthesis using Vivado 2023.2); Latency for Outer Unrolled Pipeline shows clock cycles until the first result, and in parenthesis the latency for the next immediate result in the pipeline.}
    \label{tab:mmm_unit_overhead}
\end{table}

\subsection{Tool-Aided Montgomery multipliers}
\label{extra_mmm_sec}

The next Montgomery multipliers we implemented and evaluated are the tool-aided ones: the complete HLS variant and the Karatsuba-based designs. Starting with the complete HLS design, we follow the CIOS algorithm for the HLS-C++ implementation. The compiler HLS pragmas that we used were the array partition that performs a complete array partition on the $T$ array so that we can replicate a similar behavior to the LUTRAM-style implementations of our main Montgomery multipliers, the operation binding pragma for implementing all multiplications and additions using DSPs, and the pipeline pragma for controlling the pipeline behavior of the overall design.

For the final HLS design being evaluated both the inner loops and the outer loop have the pipeline feature disabled through the specific pragma because pipelining either of the loops resulted in a complete unroll which consumed too many resources for a Montgomery Multiplier. In our opinion this kind of unit should be kept simple and compact in order to be used in more complex systems such as an ECC engine. When pipelining only the outer loop, and keeping both inner loops non-pipelined, synthesis estimations showed a LUT and FF usage of 25089 and 37621. Pipelining the outer loop and both inner loops (the entire design) showed an LUT and FF usage of 64232 and 155823. The fully pipelined design was generated with a pipeline depth of 461. These estimations are retrieved for the 32-bit word size implementation and as a comparison for the non-pipeline design we used for evaluation, LUT and FF consumption was only 11908 and 16086 after initial synthesis estimation. In all the synthesis estimations mentioned, the HLS read and write functions for providing AXI Stream access to global memory are also included alongside the main HLS Montgomery Multiplication function.

Using the non-pipeline version of the HLS implementation would also be a fair comparison to the RTL implementations, as except for the OUP design, the rest of the designs are also not pipelined. As we care for performance and obtaining high throughput we still tried to run an implementation for the full pipeline version of the HLS design (both inner and outer loops with pipeline enabled). Unfortunately, the place-and-route step could not be completed successfully because of unrouted nets caused by high congestion in the design. The design could be fully placed and routed with further optimization of the HLS code and also perhaps a different implementation strategy, but this would deviate this design from the established points and constraints of the analysis imposed over all the other designs, which is why we decided to not pursue this direction.  

The second tool-aided Montgomery multiplier approach is implemented in Verilog, but relies on the synthesizer tool to optimize the arithmetic multiplication of the words in the CIOS algorithm. We call it Karatsuba-based, from the Karatsuba multiplication~\cite{eyupoglu2015performance} which is faster than the classic schoolbook multiplication. As we wanted to see the QoR that the synthesizer can output, we also took the liberty of using this multiplication algorithm as it performs better from a latency point of view. 

The algorithm uses a divide-and-conquer approach, where the two operands involved in the multiplication are split into equal halves, which then only need three total multiplications and two extra additions compared to the four multiplications involved in the classic approach. As our DSPs can perform a square multiplication of up to 17 by 17-bits, we decided to implement only the 32 and 64-bit variants for this evaluation, so that for the 32-bit variant we only perform a single level of recursion and split the words into 16-bit and 17-bit (after additions) digits which fit into a single DSP. For the 64-bit variant, we perform a two level recursion of the Karatsuba algorithm, splitting the words first into 32-bits, and afterwards, split them again into 16-bit digits so that we can help the synthesizer better optimize the DSP usage. Equation~\ref{karatsuba} shows the general approach of implementing a one-level split and multiplication ($P = A\times B$), where the \textit{mul} function represents a classic multiplication and $n$ represents the bit-size of the inputs $A$ and $B$. All arithmetic operations are implemented through the standard operators available in the Verilog language: +/-/*.

\begin{equation} \label{karatsuba}
\begin{gathered}
A = (A1, A0) \\
B = (B1, B0) \\
Y = mul(A1 + A0, B1 + B0) \\
U = mul(A0, B0) \\
Z = mul(A1, B1) \\
P = U + (Y-U-Z)\times 2^{n/2} + Z\times 2^n \\
\end{gathered} 
\end{equation}

For DSP usage, there are two options available through Vivado's synthesizer. The first one is the automatic behaviour which is enabled by default. After writing the Verilog code, we synthesize and implement the code without any additional modifications, and the tool will try to optimize and find the best possible usage of DSPs for the provided design. We use the \textit{Flow\_PerfOptimized\_high} synthesis strategy. The second one is the usage of a synthesizer attribute called \textit{use\_dsp} which can be placed in the RTL code on signals, entities, modules, etc. Through the usage of this attribute on the Karatsuba multipliers at the module level, we instruct the synthesizer to place all the arithmetic operations involved in the module into DSPs. 

As the automatic behavior is influenced by timing constraints and selected strategies, and could also implement the arithmetic logic through fabric (LUTS, FFs) instead of DSPs if it sees fit, we evaluate both cases in order to see if default behavior and manual guidance of the synthesizer can obtain better results compared to the manual instantiations of the DSP primitives in the other Montgomery multipliers. The comparison would also show if a modern synthesizer plus a faster multiplication algorithm written naively can show better results compared to a classic and finer-grained control of the primitive resources. 

Table~\ref{tab:word_unit_karatsuba} shows the resource usage and latency of the Karatsuba multiplier Verilog implementations for 32 and 64-bits after synthesis. As mentioned above, we implemented two flavors of this approach, one where we use the default behavior and let the synthesizer pick where to infer DSPs and one where we force the usage of DSPs on the entire module by setting the \textit{use\_dsp} attribute to \textit{true}. Both implementation styles have the same latency as this is not affected by this attribute, but only by the behavioral description of the unit. 

Compared, functional-wise, to the MUL\_32 word arithmetic unit from Table~\ref{tab:word_unit_overhead} (4 DSPs, 0 LUTs, 130 FFs, 6 clock cycles latency), we can see that both 32-bit implementations have the same latency, and for the default settings (\textit{use\_dsp=auto}), it uses the same number of DSPs, but with an additional number of 85 LUTs and 102 FFs. In the 32-bit Karatsuba unit where we force the usage of DSPs on the entire module (this includes the multiplications and additions from Equation~\ref{karatsuba}), the synthesizer uses only one LUT, at the expense of 8 DSPs and a closer number of 134 FFs.

In the case of the 64-bit implementations, the MUL\_64 unit uses 16 DSPs, 422 LUTs and 473 FFs, with a latency of 18 clock cycles. The Karatsuba variants have a better latency of 11 clock cycles (this includes the 6 clock cycles from the 32-bit Karatsuba multipliers internally instantiated), but use over 1000 FFs. For the default attribute setting, we have better DSP usage (12), and a similar LUT usage (540), while forcing the attribute setting to \textit{true} uses only 8 LUTs, but increases the DSP expense at 35 primitives.

\begin{table}[!htp]
    \centering
    \begin{tabular}{c|c|c|c|c|c|c}
       Word arithmetic unit & Bit Size & DSPs & CLB LUTs & CLB Regs & Latency & Pipelined \\\hline
       Karatsuba(use\_dsp=true) & \makecell{32-bit \\ 64-bit}  & \makecell{8 \\ 35} & \makecell{1 \\ 8} & \makecell{134 \\ 1301} &  \makecell{6 \\ 11} & YES \\\hline
       Karatsuba(use\_dsp=auto) & \makecell{32-bit \\ 64-bit}  & \makecell{4 \\ 12} & \makecell{85 \\ 540} & \makecell{232 \\ 1903} &  \makecell{6 \\ 11} & YES \\\hline
    \end{tabular}
    \caption{Number of DSPs, LUTs, Registers (FFs) and latency (clock cycles) for the Karatsuba-based word arithmetic unit(post-synthesis)}
    \label{tab:word_unit_karatsuba}
\end{table}

The Karatsuba units can only handle multiplications, the remaining two additions employed in $(CARRY, T[i+j]) = A[j] \times B[i] + T[i+j] + CARRY$ are handled in the top design representing the Montgomery Multiplier. The Montgomery Multiplier module in this case is a copy of the Row-Serial design, as the Karatsuba units are pipelined and can follow a similar input operand scheduling to the one of the MADDCARRY units, but with the additions being handled separately after each multiplication result is outputted by the Karatsuba unit every clock cycle. For the Montgomery Multipliers using the Karatsuba units, we also evaluate both cases of default and forced usage of DSPs on these two chained additions. We place the attribute in this case on the wires that are used for outputting the two successive additions applied to the Karatsuba result. The two consecutive additions are naively implemented in Verilog through the addition operator in a combinational way using the \textit{assign} statement. Even though naive we wanted to see if this kind of approach can give better or similar results compared to an implementation where the arithmetic operations are optimized to better use the available resources. 

Table~\ref{tab:additional_unit_overhead} shows the resource consumption and achieved latency in clock cycles for the HLS-only and Karatsuba-based designs. Even if the Karatsuba-based designs are a lightly modified version of the Row-Serial design (replacing the word arithmetic unit with the Karatsuba units), the LUTs and FFs usage is similar to the one of the Row-Parallel designs for the 32-bit variant. Also for the 32-bit variant, letting the synthesizer choose when to use DSPs, we have 4 DSPs compared to 5 in the Row-Serial, but forcing the usage of DSPs, the entire Montgomery Multiplier ends up having 9 DSPs (almost double compared to our manually instantiated DSP design). Latency-wise, the 32-bit variant is in between the Row-Serial and Row-Parallel implementations.

For the 64-bit variant, we see an interesting fact, that the synthesizer with the forced behaviour of using DSPs, obtains a similar LUT and FF consumption compared to Row-Serial, but with almost a double usage of DSPs. The automatic behaviour, on the other size, consumes more LUTs and FFs (around 27\% for LUTs, and 43\% for FFs), but ends up using only 12 DSPs (versus the 19 needed in Row-Serial) for the entire Montgomery Multiplier. As we have a small latency of only 11 cyles for the 64-bit multiplications and we use a naive implementation of computing the extra two additions in a single cycle following the multiplier result we end up needing just 12 cycles for executing a single iteration of the inner loop of the CIOS algorithm (compared to 30 cycles needed in the 64-bit variant of Row-Serial). Thus, in total, the CIOS algorithm requires only 290 cycles (our lowest latency of all designs) when using the 64-bit variant of the Karatsuba-based Montgomery Multiplier. As latency is very important in ECC applications, this is one of the reasons we applied such a naive implementation for the two consecutive additions as we wanted to see how good the synthesis and implementation tools are in offering a compact and high-frequency design while keeping a small latency without having to worry about optimizing the FPGA primitives usage ourselves. This kind of approach in describing hardware also increases time-to-market as the designer uses the available operators to quickly implement a certain algorithm instead of having to instantiate or infer certain primitives, which can further lead to bugs that can be hard to track down in complex designs.

For the HLS implementations, the used estimations are only for the HLS-C++ function implementation that describes the CIOS algorithm, without any additional overhead of HLS wrappers used for moving data from and to the global memory, in order to have a fair comparison with the RTL implementations of the core Montgomery units. Given the disabled pipelines that should reduce resource usage, the HLS implementations still have a greater consumption compared to the Row-Serial design. It uses more than double the number of DSPs, at least three times more LUTs and almost three times more FFs. Also, the latency obtained by the tool is up to 10 times greater than the one from the Row-Serial design. As mentioned at the beginning of the subsection, we disabled the pipelines for all loops in order to reduce resource consumption and avoid any design congestion that would result in an incomplete routed design, but as we can see this comes with the cost of an increased latency induced by the HLS tool.

\begin{table}[!htp]
    \centering
    \begin{tabular}{c|c|c|c|c|c|c}
       Montogmery Multiplier & Bit Size & DSPs & CLB LUTs & CLB Regs & BRAM & Latency\\\hline
       
       HLS-Only & \makecell{24-bit \\ 32-bit \\ 64-bit}  & \makecell{8 \\ 10 \\ 42} &  \makecell{4725 \\ 4409 \\ 5041} &  \makecell{5160 \\ 5554 \\ 7267} & \makecell{0 \\ 0 \\ 0} & \makecell{9169 \\ 4021 \\ 979} \\\hline
       
       Karatsuba(use\_dsp=true) & \makecell{32-bit \\ 64-bit}  & \makecell{9 \\ 37} &  \makecell{1716 \\ 1636} &  \makecell{1832 \\ 3116} & \makecell{0 \\ 0} & \makecell{493 \\ 290} \\\hline
       
      Karatsuba(use\_dsp=auto) & \makecell{32-bit \\ 64-bit}  & \makecell{4 \\ 12} &  \makecell{1648 \\ 2192} &  \makecell{1923 \\ 3716} & \makecell{0 \\ 0} & \makecell{493 \\ 290} \\\hline
    
    \end{tabular}
    \caption{Resource consumption and latency (clock cycles) for each bit-size variant of the tool-aided Montgomery multiplier units (post-synthesis).}
    \label{tab:additional_unit_overhead}
\end{table}

\section{Results and analysis}
\label{sec:results}

We obtain and evaluate the results of the resource usage for each design together with the throughput in both Montgomery multiplications per second and MBs per second, and finally the achieved frequency. The values are retrieved for the full end-to-end design implementations, meaning the Montgomery multipliers with the attached \textit{ap\_ctrl\_chain} block-control interface, the HLS/Verilog wrapper for providing the AXI Stream interfaces and the HLS wrapper for providing the AXI4 interface used for access to the global memory. This is important, as the Montgomery multiplier needs all this overhead in order to be used and tested as we will see in the following subsection, and this overhead affects the final frequency, throughput and power usage, thus it must be mentioned as part of the whole design. We also provide the power estimations individually for the entire kernel deployed into the FPGA, but also for the core Montgomery Multiplier units, without the overhead added by the HLS wrappers (the ones described in Section~\ref{main_mmm_sec} and Section~\ref{extra_mmm_sec}). We also provide a real power consumption, measured during the benchmark of the designs, retrieved from the internal sensors available in the FPGA.

All designs are synthesized and implemented through the same strategy configuration using the Vitis 2023.2 environment, placed in the same Super Logic Region (SLR) of the FPGA, and connected to the same HBM channel ports in order to have a similar ground of testing between all designs. Listing~\ref{xclbin_config} shows the synthesis and implementation setup used in the Vitis configuration file for generating our binary file containing the hardware kernel. The settings used for implementation mimic the Performance\_ExtraTimingOpt implementation strategy but the opt\_design step is modified to \textit{Explore} instead of the default behavior as we observed this achieved better frequencies in most of the designs compared to other approaches.

\begin{lstlisting}[language=bash, caption=Vivado synthesis and implementation strategies used in evaluation of each design, captionpos=b, label=xclbin_config, breaklines=true]
[vivado]
prop=run.synth_1.STRATEGY=Flow_PerfOptimized_high
prop=run.impl_1.STRATEGY=Performance_ExtraTimingOpt
prop=run.impl_1.STEPS.OPT_DESIGN.ARGS.DIRECTIVE=Explore
prop=run.impl_1.STEPS.PLACE_DESIGN.ARGS.DIRECTIVE=ExtraTimingOpt
prop=run.impl_1.STEPS.PHYS_OPT_DESIGN.IS_ENABLED=true
prop=run.impl_1.STEPS.PHYS_OPT_DESIGN.ARGS.DIRECTIVE=Explore
prop=run.impl_1.STEPS.ROUTE_DESIGN.ARGS.DIRECTIVE=NoTimingRelaxation
\end{lstlisting}

For benchmarking the designs, we generate a set of one million random inputs and corresponding correct outputs from the NCC group implementation of the \textit{BLS12-381} Montgomery Multiplier implementations~\cite{mmm_rust}. With this generated set, we split it into five different batch size configurations which we execute 10 times each and compute the average throughput, latency and power consumption per batch, and afterward per the entire run to get the final average values. The five batch sizes are configured as: 100.000, 200.000, 250.000, 500.000 and 1.000.000 input pairs per batch. 

These batch sizes are evaluated in a single run, meaning we start with the 100.000 inputs per batch, which we send in sequence one after the other to the FPGA for 10 iterations and measure for each one the latency, throughput and power consumption. Afterward, we compute the average throughput, latency and power consumption for the 10 batches of 100.000 input pairs, save it, and move on to the next batch size of 200.000 input pairs per batch where we repeat the same process and save the average results. We repeat this for all five batch sizes mentioned before, and at the end, we compute the final throughput and power consumption of each design by computing the mean value of each average measurement obtained per batch. We also perform a warm-up phase before the actual benchmark, where we load the entire test vector of 1.000.000 inputs into the host memory and execute 100 consecutive batches of 10.000 elements.

As we use the out-of-the-box XRT and OpenCL APIs~\cite{munshi2011opencl} provided by AMD-Xilinx for communication between the host application and the FPGA, at each batch run we allocate the corresponding OpenCL buffers that will hold the batch of inputs and returned outputs. The measurement of latency, throughput and power does not include the buffer allocation but starts from the memory transfer from host to the global memory of the FPGA through the call of the \textit{enqueueMigrateMemObjects} OpenCL function, followed by the \textit{enqueueTask} function call which starts the kernel execution, and ending with another \textit{enqueueMigrateMemObjects} function call for moving the results from the FPGA back to the host application. At each batch run, we also verify the correctness (separately, not taken into measurement) of our results with the golden values obtained from the NCC group implementation. 

As an important mention for the results obtained, as the pre-computed \textit{R} for the Montgomery multiplication is $2^{384}$ which satisfies the condition $R>4\times P$, where $P$ is the \textit{BLS12-381} modulus, we don't need to implement and perform in our hardware designs the final comparison and subtraction required by the CIOS algorithm in case of modulus overflow. This does not affect the final result, as eventually, when we convert the result back from the Montgomery domain to the field domain, the conversion will correct the modulus overflow as shown and demonstrated in~\cite{walter1999montgomery, hachez2000montgomery} when $R$ is chosen to satisfy the before mentioned condition. 

We perform double verification of the returned results from the FPGA to confirm this in the benchmarking host application, first, we compare the direct result in the Montgomery domain, and in case it is different from the one computed with the Rust implementation (the final check and subtraction are implemented here), we perform the conversion to the field domain for both values and compare them again, and indeed in the few cases where we have a different result in the Montgomery domain, the field domain value is the same between the FPGA result and software implementation, thus proving the correct mathematical implementation of our hardware designs. This also allows one to perform multiple successive additions, subtractions and multiplications in the Montgomery domain, without having to perform this additional verification, as converting the end result back into the field domain will always come with the modulus overflow correction.

Tables~\ref{tab: throughput_freq} and~\ref{tab: resource_power} show the throughput, frequency, resource and power consumption of all designs implemented and benchmarked on the Avleo U55C FPGA board from AMD-Xilinx. Frequency, throughput and resource consumption are depicted for the entire end-to-end design which include the core MMM unit and the described HLS wrappers and HLS/AXI stream buffers used for streaming data from and to the global memory of the FPGA. The area consumption shown in Table~\ref{tab: throughput_freq} does not include any part of the static and/or dynamic regions introduced by the FPGA shell and Vitis tools.  

Power consumption is presented in three versions. Column $\textit{Power/run}$ shows the power consumption retrieved during benchmarking from the internal sensors of the FPGA board and besides the full hardware kernel (core MMM unit, plus HLS wrappers and stream buffers), this measurement also includes the power usage of other elements such as the global HBM memory, the XDMA sub-unit and everything else that is added automatically by Vitis in the dynamic region of the design for control and status of the hardware kernel. This value does not include the power consumption of the static region that comes pre-loaded with the Xilinx runtime shell used to deploy hardware kernels.

The \textit{Power/kernel} column represents the power usage of the hardware kernel alone(who's components are mentioned earlier), and the \textit{Power/MMM} represents the power usage of the Montgomery Multiplier core unit. These values are retrieved from the Vivado power reports, post-implementation. For the Outer Unrolled Pipeline design, the power per MMM includes the AXI Stream interface and the XPM FIFO AXIS buffers. The other RTL designs not being pipelined, do not contain the streaming interface and buffers, only the \textit{ap\_ctrl\_chain} interface for the block-control handshake mechanism needed when using the RTL Blackbox feature of Vitis.

\begin{sidewaystable}
\caption{Throughput expressed in both MBs/s and ops/s (Montgomery multiplications per second), and frequency achieved for all fully implemented, placed-and-routed designs on the AMD-Xilinx Alveo U55C FPGA board (UltrascalePlus HBM architecture).}
\begin{tabular}{c|ccccc}\toprule
\multicolumn{5}{c}{\textbf{MAIN DESIGNS}} \\\cmidrule{1-5}
\textbf{Design Name} &\textbf{Word size (bits)} &\textbf{Throughput(MBs/s)} &\textbf{Throughput ($\times10^6$ ops/s)} &\textbf{Freq (Mhz)} \\\midrule
\multirow{3}{*}{\textbf{Outer Unrolled Pipeline (OUP)}} &24 &320.3220 &7.11826 &448 \\
&32 &354.4370 &7.96803 &467 \\
&64 &228.5540 &5.07897 &486 \\\midrule
\multirow{3}{*}{\textbf{Row-Parallel (RP)}} &24 &48.2481 &1.07218 &553 \\
&32 &54.6248 &1.21388 &530 \\
&64 &53.0342 &1.17854 &526 \\\midrule
\multirow{3}{*}{\textbf{Row-Serial (RS)}} &24 &30.0159 &0.66702 &548 \\
&32 &43.1271 &0.95838 &545 \\
&64 &45.2852 &1.00634 &516 \\\midrule
\multirow{3}{*}{\textbf{Row-Serial-BRAM (RS-BRAM)}} &24 &25.3031 &0.56229 &526 \\
&32 &35.1060 &0.78013 &516 \\
&64 &40.6129 &0.90251 &523 \\\midrule
\multicolumn{5}{c}{\textbf{ADDITIONAL DESIGNS}} \\\cmidrule{1-5}
\textbf{Design Name} &\textbf{Word size (bits)} &\textbf{Throughput (MBs/s)} &\textbf{Throughput ($\times10^6$ ops/s)} &\textbf{Freq (Mhz)} \\\midrule
\multirow{3}{*}{\textbf{HLS-only}} &24 &1.9486 &0.04330 &514 \\
&32 &4.4035 &0.09782 &510 \\
&64 &13.8468 &0.30771 &422 \\\midrule
\multirow{2}{*}{\textbf{Karatsuba-based (use\_dsp=true)}} &32 &25.8984 &0.57552 &290 \\
&64 &45.7331 &1.01629 &305 \\\midrule
\multirow{2}{*}{\textbf{Karatsuba-based (use\_dsp=auto)}} &32 &28.9897 &0.64422 &328 \\
&64 &54.5536 &1.21230 &368 \\
\bottomrule
\end{tabular}
\label{tab: throughput_freq}
\end{sidewaystable}

\begin{sidewaystable}
\caption{Resource and power consumption for all fully implemented, placed-and-routed designs on the AMD-Xilinx Alveo U55C FPGA board (UltrascalePlus HBM architecture).}
\begin{tabular}{c|ccccccccc}\toprule
\multicolumn{9}{c}{\textbf{MAIN DESIGNS}} \\\cmidrule{1-9}
\textbf{Design Name} &\makecell{\textbf{Word size} \\ \textbf{(bits)}} & \makecell{\textbf{Power/run} \\ \textbf{(W)}} &\makecell{\textbf{Power/MMM} \\ \textbf{(W)}} &\makecell{\textbf{Power/kernel} \\ \textbf{(W)}} &\textbf{LUTs} &\textbf{DSPs} &\textbf{FFs} &\textbf{BRAMs} \\\midrule
\multirow{3}{*}{\textbf{Outer Unrolled Pipeline (OUP)}} &24 &6.362 &1.587 &2.428 &18850 &48 &37761 &23 \\
&32 &6.155 &1.321 &2.165 &16295 &60 &40944 &23 \\
&64 &6.237 &1.598 &2.445 &15308 &114 &31084 &23 \\\midrule
\multirow{3}{*}{\textbf{Row-Parallel (RP)}} &24 &3.568 &0.204 &1.247 &10894 &42 &12405 &21 \\
&32 &3.558 &0.304 &1.272 &9547 &59 &13191 &21 \\
&64 &6.362 &0.588 &1.760 &10924 &120 &14641 &54 \\\midrule
\multirow{3}{*}{\textbf{Row-Serial (RS)}} &24 &3.570 &0.065 &0.923 &7783 &3 &11170 &21 \\
&32 &3.455 &0.079 &0.930 &7144 &5 &11204 &21 \\
&64 &6.179 &0.133 &1.387 &8146 &19 &11456 &54 \\\midrule
\multirow{3}{*}{\textbf{Row-Serial-BRAM (RS-BRAM)}} &24 &3.557 &0.054 &0.952 &7374 &3 &10944 &22 \\
&32 &6.166 &0.074 &1.090 &7100 &5 &10384 &55 \\
&64 &6.202 &0.135 &1.176 &7585 &19 &11093 &55 \\\midrule
\multicolumn{9}{c}{\textbf{ADDITIONAL DESIGNS}} \\\cmidrule{1-9}
\textbf{Design Name} &\makecell{\textbf{Word size} \\ \textbf{(bits)}} & \makecell{\textbf{Power/run} \\ \textbf{(W)}} &\makecell{\textbf{Power/MMM} \\ \textbf{(W)}} &\makecell{\textbf{Power/kernel} \\ \textbf{(W)}} &\textbf{LUTs} &\textbf{DSPs} &\textbf{FFs} &\textbf{BRAMs} \\\midrule
\multirow{3}{*}{\textbf{HLS-only}} &24 &6.185 &0.357 &1.821 &11742 &8 &14750 &54 \\
&32 &6.159 &0.403 &1.829 &11605 &10 &14945 &54 \\
&64 &6.062 &0.557 &2.051 &12249 &42 &16500 &54 \\\midrule
\multirow{2}{*}{\textbf{Karatsuba-based (use\_dsp=true)}} &32 &3.351 &0.067 &0.683 &7479 &9 &11096 &21 \\
&64 &3.414 &0.336 &0.906 &7515 &37 &12043 &21 \\\midrule
\multirow{2}{*}{\textbf{Karatsuba-based (use\_dsp=auto)}} &32 &3.286 &0.068 &0.729 &7812 &4 &11187 &21 \\
&64 &3.441 &0.273 &0.863 &8185 &12 &12494 &21 \\
\bottomrule
\end{tabular}
\label{tab: resource_power}
\end{sidewaystable}

As expected, the fastest designs from the throughput point of view are the Outer Unrolled Pipeline (OUP) ones, thanks to the small latency of 52/48/84 clock cycles (for the 24/32/64-bit word sizes) for outputting a new result. Frequency-wise, Row-Serial for 24/32-bits obtained some of the highest values, but the best one was obtained for a Row-Parallel design, the 24-bit version, clocking 553MHz. The HLS 24/32-bit versions also achieved over 500MHz, while the Karatsuba-based implementations obtained the lowest values in the list, between 290MHz and 368MHz.

Power per MMM unit (obtained post-implementation) and area-wise per kernel, the Row-Serial designs obtained the best results in terms of a compact and low-power design, but the Karatsuba-based designs on 32-bits obtained power values per MMM unit close to Row-Serial on 24-bits, and better than Row-Serial on 32-bits. When looking at Power per kernel (also obtained post-implementation), the Karatsuba-based designs have the lowest values of all designs, but this is also influenced by the lower frequency achieved which influences the final power usage. Suppose a power evaluation between the actual design choices is desired. In that case, the Power per MMM unit is more suitable for this purpose, as this value returned by Vivado evaluates the power usage of the isolated MMM design composed of the power consumption of the resources, clocks, and signals that are part and used in the multiplier alone. The Power per Kernel takes into consideration also the HLS wrappers, which have the overhead of the resources themselves, but also accumulate the clocks and signals power draw that go into the power consumption of the entire kernel.

An interesting behavior of the Vitis linker tool that implements the entire kernel is the usage of BRAMs which is not always constant between implementations. The BRAMs are used in the implementation of the read\_input/mem2stream and write\_result/stream2mem HLS wrappers for moving data between the global memory and the MMM units, but even though we used the same HLS code across all designs, the tool used a different number of blocks between designs, especially as the wrappers operate only on a full 512-bit bandwidth for moving the 384-bit operands inside the MMM unit, so it should not be impacted by the word size used in the implementations. This has a direct correlation to the Power per run, as except for the Outer Unrolled Pipeline designs, all other designs that used 54-55 BRAMs had a Power per run consumption of over 6W, while the other designs that used 21-22 BRAMs, had almost half of that power consumption. The BRAM-based designs have an extra BRAM used directly in the RTL description for the register file that holds the partial products.

\begin{figure}[!htp]
\centering
    \includegraphics[width=1\textwidth]{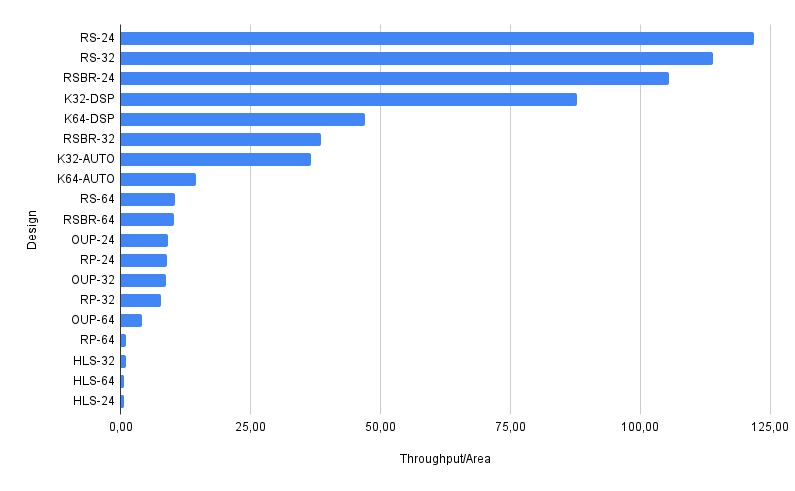}
    \caption{Throughput to Area efficiency (higher is better) ranking of each design. The throughput used is ops/s, and Area is calculated as $LUTs\times DSPs \times FFs \times BRAMs$. Area resources used are from Table~\ref{tab: resource_power} which depict the resources for implementing the full hardware kernel.}
    \label{fig:throughput-area}
\end{figure}

\begin{figure}[!htp]
\centering
    \includegraphics[width=1\textwidth]{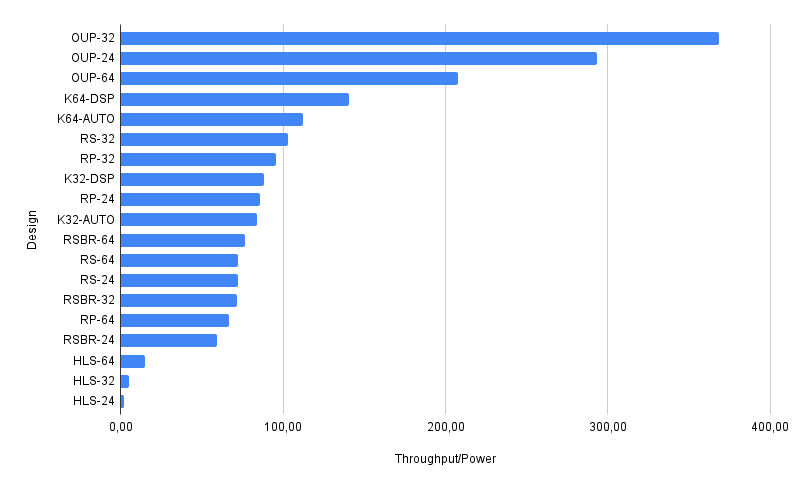}
    \caption{Throughput to Power efficiency (higher is better) ranking of each design. The throughput used is ops/s, and the Power used is from Table~\ref{tab: resource_power}, column \textit{Power/kernel}, which depicts the power usage of the full hardware kernel.}
    \label{fig:throughput-power}
\end{figure}

Figures~\ref{fig:throughput-area} and~\ref{fig:throughput-power} show the efficiency ranking from the highest to lowest for area-efficiency and power-efficiency. \textit{RSBR} represents the Row-Serial design with BRAM implementation, \textit{K32/K64-AUTO} represents the Karatsuba-based designs with the default/automatic behaviour of using DSPs, and \textit{K32/K64-DSP} represents the Karatsuba-based designs with the forced attribute of using DSPs. Throughput used is Montogmery multiplications per second, area is computed as the product between LUTs, DSPs, FFs, and BRAMs from Table~\ref{tab: resource_power} for the full hardware kernel, and power used is also for the kernel implementation (column \textit{Power/kernel} from the same table).

Even if the power consumption on any level (MMM unit, kernel, run) recorded by the OUP designs was the highest one, thanks to the much higher throughput compared to the other designs, the overall power efficiency of these designs is the best among all, with the 32-bit variant being the fastest and most efficient power-wise. The Karatsuba-based designs implemented on 64-bits follow afterward, as these designs also have a better throughput compared to all other non-pipeline designs thanks to the small latency of only 290 cycles.

As area efficiency, the Row-Serial designs (24 and 32-bits) score the best values, thanks to the compact design across all types of resources, followed by the 32-bit Row-Serial BRAM-based implementation, and again as part of the top five, by the Karatsuba-based implementations. Even though the default behavior of DSP usage for the Karatsuba implementation on 64-bit scored the best throughput among the non-pipeline designs, power/area-efficiency was better overall for the version where we forced the usage of DSPs.

Regarding comparison to the software implementation of the Montgomery multiplication presented in \cite{mmm_rust}, we've run their provided benchmark unit tests on an Intel(R) Xeon(R) CPU E5-2630 v4 CPU running at 2.20GHz, with 64GB of RAM available. The author describes multiple variants of the implementation, the most notable ones being a pure Rust implementation, a Rust implementation using the available intrinsic, and an optimized assembly routine implementation for the Montgomery multiplication. On our machine, the average execution times for a single multiplication were 62.09ns for pure Rust, 53.75ns for Rust with intrinsic, and 50.93ns for the assembly implementation. This translates to a throughput of 898.80 MBps, much higher than the FPGA implementations. 

Because area cannot be used for efficiency in the case of the CPU, we measured the peak power usage of the CPU during the application run, which went from an average of 14.69 W in idle to 44.38 W during the actual multiplication run. This translates to a power usage of 29.69 W during the multiplication run, offering a throughput (MBps) to power (W) efficiency of 30.27. In comparison, using our full kernel run power draw of 6.155W for the OUP-32 design, our throughput to power efficiency is 57.58, almost two times better compared to the CPU implementation. For FPGA, we used the throughput from Table~\ref{tab: throughput_freq} where we benchmarked the average throughput for the entire hardware kernel, which also includes the data transfer between the host and the global memory of the FPGA, and further the data streaming between the global memory and the core MMM unit.

As a mentioned, the power per run measured for the FPGA is also the difference between the peak power usage during the run, and the power usage of the board in idle (no hardware kernel running). This offers a fair comparison to the CPU, where we applied a similar approach. This offers a better view of the power overhead the application or hardware kernel adds on top of the devices running in idle mode. Also, the average idle power consumption of the FPGA is around 14.77W, very close to the average power of the CPU in idle (14.69 W).

\section{Limitations and future work}

From our perspective, all our designs could be further improved and optimized to increase performance and efficiency. For example, in the computing of the real application throughput, we use the FPGA shell provided out-of-the-box by AMD-Xilinx and the logic overhead added in the dynamic region by the Vitis tool when performing the linking phase of the implementation. This comes with a bottleneck in terms of achieved frequency, as the tool inserts different monitors and control logic for watching and manipulating the hardware kernels from the user application. We observe this even during implementation, as many of the messages regarding the critical path of the design were reported by the tool on the dynamic region logic, and not the actual wires or registers implemented in our multipliers. A better approach to increase frequency would be for the hardware designer to implement its own smaller version of the FPGA shell, to reduce overhead and thus the bottleneck.  

Our pipeline designs were designed to unroll the outer loop of the CIOS algorithm but keep the inner loops tightly coupled, thus a new input could not be processed until the current one does not complete both inner loops. Splitting our pipeline stages even further so that we could process each of the inner loops of an outer loop iteration individually, would greatly increase our throughput by a theoretical factor of two and could lead to around 700MBps, but at the same time would mean an increase of at least $2\times$ in the used resources. Also, this does not take in consideration the congestion that the deeper pipeline would bring, and how this could affect the critical path and final frequency.

Of course an important limitation from our point of view is in the DSP cores themselves. Having constraints such as the fixed word size and fixed position inside the FPGA fabric impacts the designs, causing an increased cost of both latencies and routing. From this point of view, trying to use only DSP units for the arithmetic operations of big-int numbers saw a higher throughput only in the OUP designs when compared to the Karatsuba-based designs. If we take into consideration area and power we can see that the throughput benefit is shadowed by the area usage and power consumption could play a high role in the requirements of the constraints as OUP has the highest consumption.

As future work, we could approach pipelining the Karatsuba-based designs and see how they compare to the OUP designs, as in the non-pipeline approach, the throughput obtained with the Karatsuba-based designs was better at a lower frequency. Of course, being a naive implementation, this could be further improved, such as using a 3:2 compressor for adding three inputs implemented in multiple stages instead of our simple approach of adding three 32/64-bit words in one cycle. Also, even through we divided the Karatsuba layers down to 16/17-bits so that the multiplication could fit into one DSP, we used a single cycle for performing it in order to have a small overall latency. One additional optimization would be to describe the multiplication in three cycles so that we could use the maximum theoretical frequency of the DSP. These optimizations would bring a small latency increase, but a higher frequency increase as well, and could overall show a better throughput.

As another future work perspective, one could use for the addition operations employed in the Montgoemry algorithm steps the CARRY8 primitives, as this allow fast additions on number up to 32-bits, and use the DSPs only for multiplications. This could relax the routing process, allowing for smaller delays and even higher frequencies, compared to the our current approach, where all arithmetic operations were performed through the DSPs.

\section{Remarks and conclusion}
\label{sec:conclusion}

In this work, we evaluated how different design choices varying from the manual instantiation of DSP primitives, to relying on the synthesis tool or using the High-Level Synthesis framework impact the performance and efficiency in the implementation of a Montgomery Modular Multiplier used for a newer elliptic curve, the \textit{BLS12-381}, which already started to be widely adopted in modern blockchain applications.

One interesting output of this case study was seeing that manual instantiation of FPGA primitives (in this case DSPs) can be avoided in the case of a performance-oriented design, as we saw that going for a more efficient Karatsuba algorithm for word multiplication gives better overall throughput while also relying on the tool to decide where and how to use DSPs. This comes in contrast to the HLS implementations, where although we were able through the compiler directives to decrease the resource usage, this came with the penalty of high latency, up to 10 times greater than our RTL Verilog designs. Interestingly even though the HLS CIOS function description had a higher resource consumption, in the final hardware kernel used for the real application test case, the LUT and FF consumption was similar overall with the other designs. 

The HLS compiler improved over the years from our perspective but still has a long path ahead its way if it wants to dominate the hardware design market. Particularly for the Vitis HLS compiler that we analyzed, we saw that through careful usage of the pragmas a hardware designer can get a compact design or a performance-oriented one but not both. At least for this use case where big integers are involved, simply implementing the C++ algorithm as one would do in the software counterpart and inserting the compiler pragmas doesn't cut. The algorithm must be (sometimes heavily) modified to guide the compiler into synthesizing efficient hardware out of it. For fast prototyping and space-time exploration of different design choices, it offers superior productivity over the RTL approach, even if we take into consideration modifying the C++ algorithm to get better results.

The Vitis HLS framework is far richer nowadays than it used to be. Hardware designers have now a plethora of tailored libraries for efficient implementation of popular algorithms into FPGA hardware kernels and it offers great integration with current FPGA boards from the AMD portfolio. The hardware-software co-simulation also improved, allowing engineers to test their design from end-to-end and the RTL blackbox feature can be the key that can offer the best of both HLS and RTL. We saw how using HLS for memory interfacing while using Verilog for the critical path modules offers the best overall design.

The Vivado synthesizer also improved compared to past versions. We saw that the synthesizer can better help a hardware designer today by taking care of optimizing the resource usage while the hardware designer can focus on writing expressive descriptions of algorithms instead of having to tweak and optimize each FPGA primitive which in the end decreases the readability of the code and can lead to errors and bugs.

Overall the entire Vitis Unified Platform seems more mature. It better helps an engineer in each step of a complex project that can rely on both HLS and RTL to implement an FPGA design. Hardware designers can use HLS for the initial exploration of ideas and later optimize the critical components through Verilog and can make use of shell script languages to automate most of the design process. Hardware verification is also much better and faster through C++ as third-party software libraries can be easily added to check the correctness of the design.

\bibliography{sn-bibliography}

\end{document}